\documentclass{aa}  
\usepackage{booktabs}
\usepackage{natbib,twoopt}
\usepackage[breaklinks=true, colorlinks=true, allcolors=blue]{hyperref} %% \bibpunct{(}{)}{;}{a}{}{,} %% natbib format for A&A and ApJ
\makeatletter
\newcommandtwoopt{\citeads}[3][][]{\href{http://adsabs.harvard.edu/abs/#3}%
{\def\hyper@linkstart##1##2{}%
\let\hyper@linkend\@empty\citealp[#1][#2]{#3}}}
\newcommandtwoopt{\citepads}[3][][]{\href{http://adsabs.harvard.edu/abs/#3}%
{\def\hyper@linkstart##1##2{}%
\let\hyper@linkend\@empty\citep[#1][#2]{#3}}}
\newcommandtwoopt{\citetads}[3][][]{\href{http://adsabs.harvard.edu/abs/#3}%
{\def\hyper@linkstart##1##2{}%
\let\hyper@linkend\@empty\citet[#1][#2]{#3}}}
\newcommandtwoopt{\citeyearads}[3][][]%
{\href{http://adsabs.harvard.edu/abs/#3}
{\def\hyper@linkstart##1##2{}%
\let\hyper@linkend\@empty\citeyear[#1][#2]{#3}}}
\makeatother

\usepackage{etoolbox}
\makeatletter
\patchcmd\@combinedblfloats{\box\@outputbox}{\unvbox\@outputbox}{}{%
   \errmessage{\noexpand\@combinedblfloats could not be patched}%
}%
 \makeatother
\usepackage{graphicx,amsmath}
\usepackage{txfonts}
\begin{document} 

   \title{Multiphase quasar-driven outflows in PG 1114+445}

   \subtitle{I. Entrained ultra-fast outflows}

   \author{Roberto Serafinelli
          \inst{1,2,3}\fnmsep\thanks{{roberto.serafinelli@inaf.it}}
          \and
          Francesco Tombesi\inst{1,4,5,6}
          \and
          Fausto Vagnetti\inst{1}
          \and
          Enrico Piconcelli\inst{6}
          \and
          \\Massimo Gaspari\inst{7}\fnmsep\thanks{{\it Lyman Spitzer Jr.} Fellow}
          \and
          Francesco G. Saturni\inst{6,8}
          }

   \institute{Dipartimento di Fisica, Universit\`a di Roma ``Tor Vergata'', Via della Ricerca Scientifica 1, 00133, Roma, Italy
         \and
         Dipartimento di Fisica, ``Sapienza'' Universit\`a di Roma, Piazzale Aldo Moro 5, 00185, Roma, Italy
         \and
         INAF - Osservatorio Astronomico di Brera, Via Brera 28, 20121, Milano, Italy
         \and
         X-ray Astrophysics Laboratory, NASA/Goddard Space Flight Center, Greenbelt, MD, 20771, USA
         \and
         Department of Astronomy, University of Maryland, College Park, MD, 20742, USA
         \and
         INAF - Osservatorio Astronomico di Roma, via Frascati 33, 00044, Monte Porzio Catone (Roma), Italy
         \and
         Department of Astrophysical Sciences, Princeton University, 4 Ivy Lane, Princeton, NJ, 08544-1001, USA
         \and
         Space Science Data Center, Agenzia Spaziale Italiana, Via del Politecnico snc, 00133, Roma, Italy}

   \date{Received XXX; accepted YYY}

% \abstract{}{}{}{}{} 
% 5 {} token are mandatory
 
  \abstract
  % context heading (optional)
  % {} leave it empty if necessary  
%   {To investigate the physical nature of the `nuc\-leated instability' of   proto giant planets, the stability of layers   in static, radiative gas spheres is analysed on the basis of Baker's   standard one-zone model.}
  % aims heading (mandatory)
%   {It is shown that stability   depends only upon the equations of state, the opacities and the local   thermodynamic state in the layer. Stability and instability can   therefore be expressed in the form of stability equations of state   which are universal for a given composition.}
  % methods heading (mandatory)
%   {The stability equations of state are  calculated for solar composition and are displayed in the domain   $-14 \leq \lg \rho / \mathrm{[g\, cm^{-3}]} \leq 0 $,   $ 8.8 \leq \lg e / \mathrm{[erg\, g^{-1}]} \leq 17.7$. These displays   may be   used to determine the one-zone stability of layers in stellar   or planetary structure models by directly reading off the value of   the stability equations for the thermodynamic state of these layers,   specified   by state quantities as density $\rho$, temperature $T$ or   specific internal energy $e$.   Regions of instability in the $(\rho,e)$-plane are described   and related to the underlying microphysical processes.}
  % results heading (mandatory)
%   {Vibrational instability is found to be a common phenomenon   at temperatures lower than the second He ionisation   zone. The $\kappa$-mechanism is widespread under `cool'   conditions.}
  % conclusions heading (optional), leave it empty if necessary 
%   {}
{Substantial evidence in the last few decades suggests that outflows from supermassive black holes (SMBH) may play a significant role in the evolution of galaxies. These outflows, powered by active galactic nuclei (AGN), are thought to be the fundamental mechanism by which the SMBH transfers a significant fraction of its accretion energy to the surrounding environment. Large-scale outflows known as warm absorbers (WA) and fast disk winds known as ultra-fast outflows (UFO) are commonly found in the spectra of many Seyfert galaxies and quasars, and a correlation has been suggested between them. Recent detections of low ionization and low column density outflows, but with a high velocity comparable to UFOs, challenge such initial possible correlations. Observations of UFOs in AGN indicate that their energetics may be enough to have an impact on the interstellar medium (ISM). However, observational evidence of the interaction between the inner high-ionization outflow and the ISM is still missing. We present here the spectral analysis of 12 XMM-Newton/EPIC archival observations of the quasar PG 1114+445, aimed at studying the complex outflowing nature of its absorbers. Our analysis revealed the presence of three absorbing structures. We find a WA with velocity $v\sim530$ km s$^{-1}$, ionization $\log\xi/\text{erg cm s}^{-1}\sim0.35,$ and column density $\log N_\text{H}/\text{cm}^{-2}\sim22$, and a UFO with $v_\text{out}\sim0.145c$, $\log\xi/\text{erg cm s}^{-1}\sim4$, and $\log N_\text{H}/\text{cm}^{-2}\sim23$. We also find an additional absorber in the soft X-rays ($E<2$ keV) with velocity comparable to that of the UFO ($v_\text{out}\sim0.120c$), but ionization ($\log\xi/\text{erg cm s}^{-1}\sim0.5$) and column density ($\log N_\text{H}/\text{cm}^{-2}\sim21.5$) comparable with those of the WA. The ionization, velocity, and variability of the three absorbers indicate an origin in a multiphase and multiscale outflow, consistent with entrainment of the clumpy ISM by an inner UFO moving at $\sim15\%$ the speed of light, producing an entrained ultra-fast outflow (E-UFO).}

%{We present here the spectral analysis of 12 XMM-Newton/EPIC archival observations of the type-1 quasar PG 1114+445.}

   \maketitle
%
%-------------------------------------------------------------------

\section{Introduction}
\label{sec:intro}
Supermassive black holes (SMBH) in active galactic nuclei (AGN) are thought to be fundamental players in the evolution of their host galaxies. This is evident since host-galaxy properties, such as the stellar velocity dispersion \citep[e.g., ][]{ferrarese00} and the mass of the bulge \citep[e.g., ][]{haring04}, are correlated with the black hole mass. This suggests the existence of a ``feedback'' mechanism between AGN activity and the star formation process, and disk winds are the most promising candidates to drive such processes \citep[e.g., ][]{king15,fiore17}.\\
\indent Moderately ionized winds ($\xi\lesssim100$ erg cm s$^{-1}$) are often detected in the soft X-ray band as blue-shifted absorption lines and edges, indicating that these so-called warm absorbers (WA) have velocities in the range $v\sim100-1000$ km s$^{-1}$ \citep[e.g., ][]{halpern84,blustin05,kaastra14}. Another class of winds, detected through X-ray absorption in AGNs, known as ultra-fast outflows (UFOs), is characterized by a much higher ionization state ($\xi\sim10^{3}-10^{6}$ erg cm s$^{-1}$) and velocity ($v\sim0.1-0.4c$), measured through absorption lines of highly ionized gas, most often Fe {\footnotesize XXV} and Fe {\footnotesize XXVI} \citep[e.g.,][]{chartas02,pounds03a, pounds03b, tombesi10a, tombesi10b, tombesi11, giustini11, gofford13, tombesi14, tombesi15, nardini15, vignali15, braito18}.\\
\indent \citet{tombesi13} found that WA and UFO parameters show overall trends in a sample of 35 Seyfert galaxies and quasars. This may suggest that WAs and UFOs could be different phases of a large-scale outflow. In particular, UFOs are launched at relativistic velocities, probably from the inner parts of the accretion disk surrounding the central SMBH, while WAs are most likely located at larger distances. However, there have been recent detections of fast outflows also in the soft X-ray band \citep[e.g.,][]{gupta13, gupta15, longinotti15, pounds16,reeves16}, with identification of highly blue-shifted K$\alpha$ lines of O {\footnotesize VI} and O {\footnotesize VII}, suggesting high velocity ($v\sim0.1-0.2c$), but lower ionization states than typical UFOs.\\
\indent UFOs are thought to have a significant impact on the interstellar medium (ISM) of their host galaxy. In fact, current models \citep[e.g.,][]{king03,king05} predict that inner UFOs may shock the ISM, transferring their kinetic energy to the ambient medium and possibly driving feedback on their host galaxy. In the aftermath of the shock, such models predict four regions being formed: (i) the inner UFO, (ii) the shocked UFO, (iii) the shocked swept-up ISM, and (iv) the outer ambient medium, not yet affected by the inner outflows. If the hot shocked gas cools down effectively, which means that the cooling time is shorter than the flow time, only momentum is conserved. In the opposite case of negligible cooling, the energy is conserved and hence the UFO transfers its kinetic power to the ISM \citep[e.g.,][]{costa14,king15}, possibly clearing out the galaxy of its own gas \citep[e.g.,][]{zubovas12}. Even though direct evidence of the shocked ISM has been claimed in the past \citep{pounds11}, it has been recently proposed that the observed low-ionization UFOs are due to shocked ISM by the inner UFO \citep{sanfrutos18}.\\
\indent In addition, \citet{gaspari17} have shown that galaxy evolution might be regulated by a duty cycle between such multiphase AGN feedback, and a feeding phase that is thought to be due to so-called chaotic cold accretion \citep[CCA,][]{gaspari13}, which is the cooling of gas clumps and clouds that ``rain'' toward the innermost regions of the AGN \citep[e.g., ][]{gaspari18}. While observational evidence of the feedback phase has been extensively collected in the last two decades in many spectral rest-frame bands, such as X-rays, optical, UV, and sub-millimetric \citep[see, e.g.,][for recent reviews]{fiore17,cicone18}, only recent observations are starting to detect the  feeding phase \citep[e.g.,][]{tremblay16,lakhchaura18,tremblay18,temi18}.\\

\indent PG 1114+445 is a type-1 quasar at $z=0.144$ \citep{hewett10}. The mass of the central SMBH and the bolometric luminosity are estimated \citep{shen11} as $\log (M/M_\odot)\sim8.8$ and $\log (L_\text{bol}/\text{erg s}^{-1})\sim45.7$, respectively, implying an Eddington ratio of $L/L_\text{Edd}\sim7\%$. The source was observed in 1996 by the Advanced Satellite for Cosmology and Astrophysics (ASCA) in the X-ray band and by Hubble Space Telescope (HST) Faint Object Spectrograph (FOS) camera in the UV band. The X-ray analysis \citep{george97} showed the presence of an ionized WA in the soft band, interpreted as photoelectric absorption edges of O {\footnotesize VII} and O {\footnotesize VIII}. Moreover, a detection of a possible absorption line at $E=7.25^{+0.42}_{-0.48}$ keV was interpreted as being due to highly ionized iron, possibly suggesting the presence of a mildly relativistic outflow with $v\sim0.1c$. The investigation of the UV spectrum \citep{mathur98} discovered C {\footnotesize IV} and Ly$\alpha$ narrow absorption lines (NAL), with line-of-sight velocities of $v_\text{out}\sim530\text{ km s}^{-1}$. Given the similar ionization state of the soft X-ray and UV absorption features, \cite{mathur98} concluded that they are likely to originate from the same material.\\
\indent In 2002, a much higher quality X-ray spectrum was obtained by a $44$ks XMM-Newton observation. These data revealed the complex two-component nature of the WA \citep{ashton04, piconcelli05}. In a recent ensemble work on a sample of optically selected quasars in XMM-Newton archival data \citep{serafinelli17}, we used 11 additional archival spectra, also included in a recent study based on the two-corona model \citep{petrucci18}. Here we present a detailed spectral analysis of these data, together with a re-analysis of both the 2002 XMM-Newton and the 1996 ASCA observations, aimed at unveiling the multiphase nature of the outflows of PG 1114+445.\\
\indent The article is structured as follows. Section~\ref{sec:datared} describes the data reduction techniques used for these spectra. We show the spectral analysis and results in Sect.~\ref{sec:spectral}. The distances of the absorbers are computed in Sect.~\ref{sec:dist}, while the energetics of the wind is described in Sect.~\ref{sec:winden}. We summarize and discuss the results in Sect.~\ref{sec:discussion}. Throughout the paper we have use the following cosmology: $H_0=70$ km s$^{-1}$ Mpc$^{-1}$, $\Omega_\text{m}=0.3$ and $\Omega_\Lambda=0.7$.

\section{Observations and data reduction} 
\label{sec:datared}

XMM-Newton observed PG 1114+445 on twelve occasions between 2002 and 2010. In May 2002 it was observed for $\sim44$ks (OBSID 0109080801). Then, a campaign of 11 observations was performed between 2010 May 19 and 2010 December 12 (sequential OBSIDs from 0651330101 to 0651331101), for a total duration of $\sim380$ ks. Details on duration and exposure of the single observations are given in Table~\ref{tab:sample}. We extracted the event lists of the European Photon Imaging Camera (EPIC) detectors, both pn and Metal Oxide Semi-conductor (MOS), with the standard System Analysis Software ({\footnotesize SAS}, version 16.0.0) tools {\tt epproc} and {\tt emproc}. All observations are affected by background particle flaring \citep[e.g.,][]{deluca04,marelli17} to different extents, particularly OBSIDs from 0651330201 to 0651330501 (see Table~\ref{tab:sample} for details). Therefore, we applied an appropriate filtering to remove the times affected by this effect. After checking that no pile-up \citep[e.g.,][]{ballet99} correction was needed for any of these observations, we extracted the spectra by selecting a region in the CCD image of $40''$ radius around the source, and the background by extracting a source-free region of the same size. We generated response matrices and auxiliary response files using the {\footnotesize SAS} tools {\tt rmfgen} and {\tt arfgen} respectively. Finally, we grouped the spectra by allowing 50 counts for each spectral bin using {\tt specgroup}, considering a minimum energy width of one fifth of the full width half-maximum (FWHM) resolution. XMM-Newton MOS1 and MOS2 spectra were combined to obtain a higher signal-to-noise ratio (S/N). We considered the $0.3-10$ keV band of each spectrum.\\
\indent We reduced the Reflection Grating Spectrometer (RGS) spectra by using {\tt rgsproc}, screening times with high particle background through the examination of the RGS light curves, and using {\tt rgsfilter} and {\tt rgsspectrum} to produce clean spectra. However, we could not perform a statistically meaningful analysis because, even combining all 24 RGS1 and RGS2 spectra together with {\tt rgscombine}, we obtained a combined mean spectrum with insufficient S/N. We also analyzed the 150 ks 1996 ASCA observation (ID 7407200\footnote{\url{https://heasarc.gsfc.nasa.gov/FTP/asca/data/tartarus/products/74072000/74072000\_gsfc.html}}), for which we retrieved the already reduced data products from the Tartarus ASCA AGN database \citep{turner01}.\\
\indent For simplicity, we gave each observation a simple identification code, where Obs.~A represents the ASCA observation, Obs.~0 is the 2002 XMM-Newton observation, while Obs.~1 to Obs.~11 label the 2010 campaign observations (see Table~\ref{tab:sample}). Finally, we grouped together Obs.~1 to 3, 4 to 5, and 8 to 9 in order to increase the count rates of the single observations, mostly due to severe background particle screening, in order to reach a level of S/N adequate for the spectral analysis.\\

\section{Spectral analysis}
\label{sec:spectral}

\subsection{Models}

\begin{figure}
\centering
\includegraphics[width=\columnwidth]{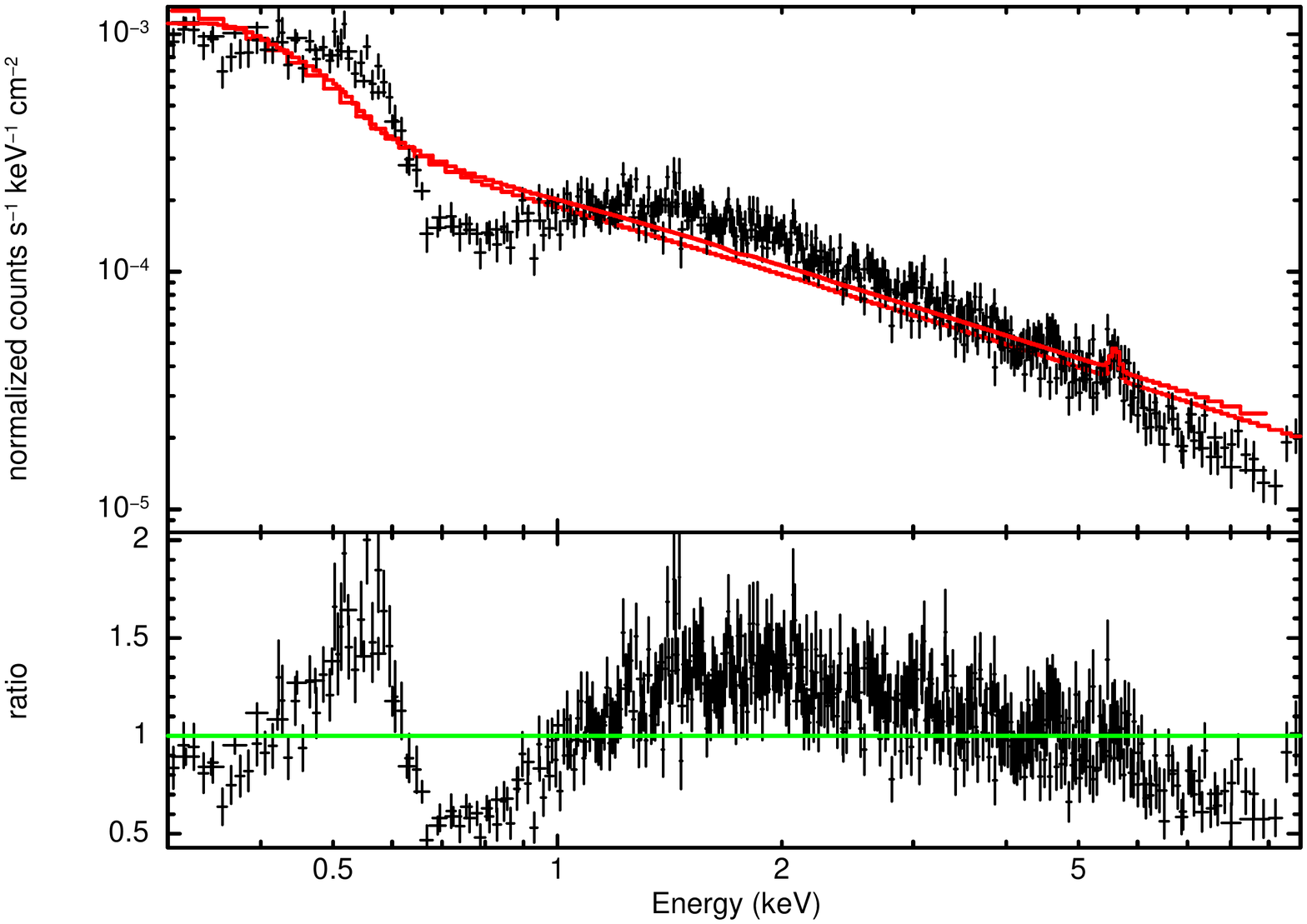}\\
\includegraphics[width=\columnwidth]{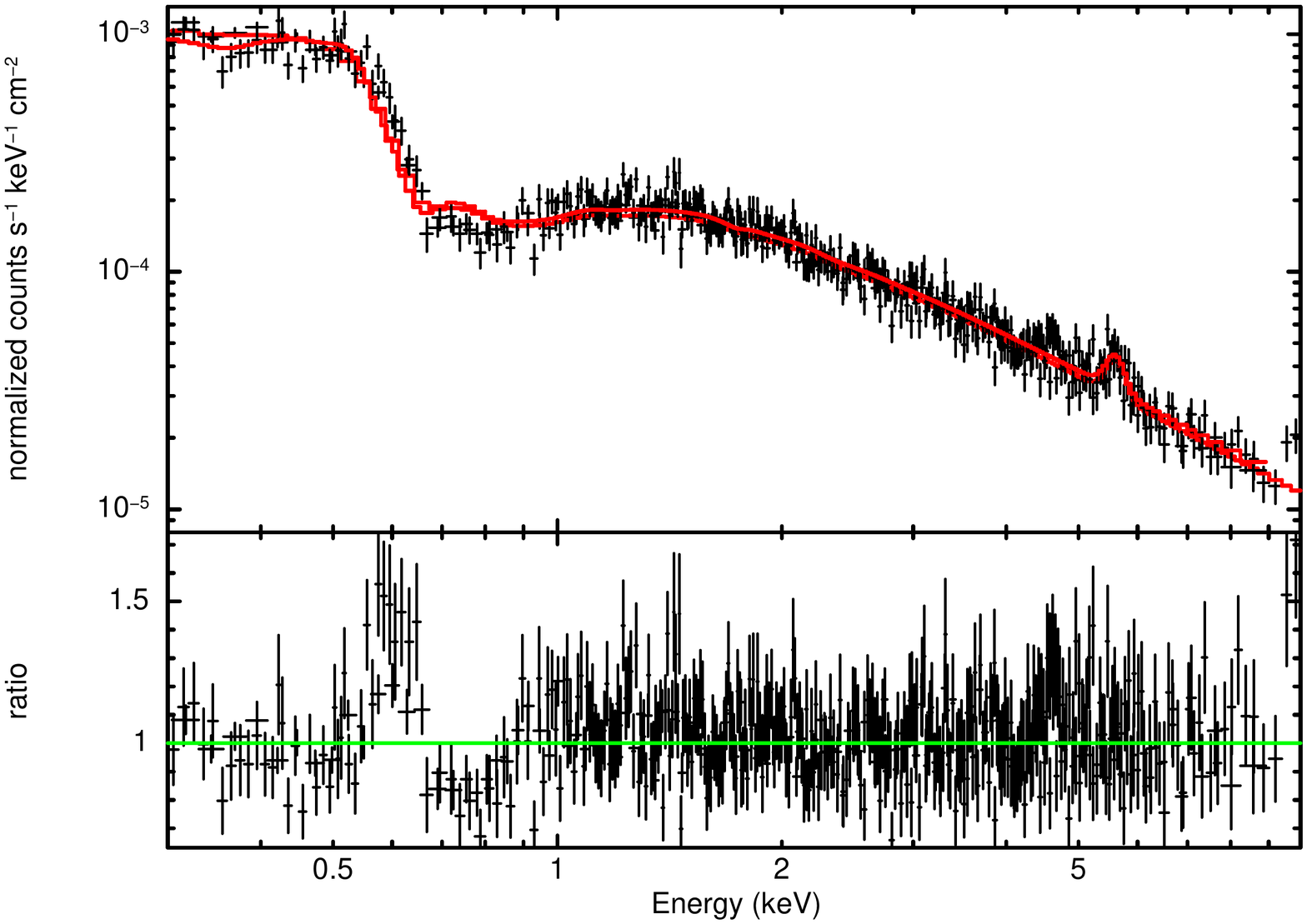}\\
\includegraphics[width=\columnwidth]{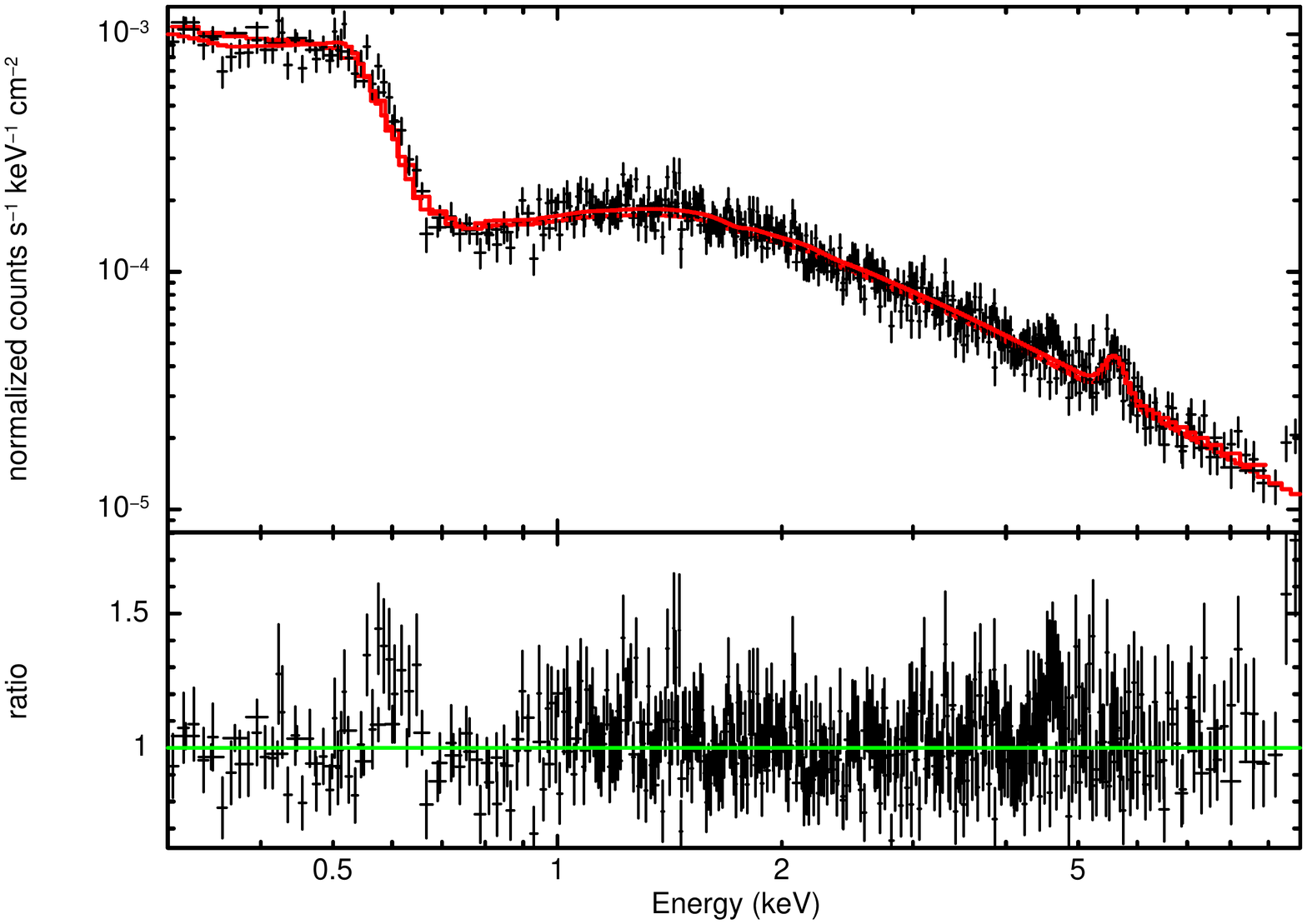}
\caption{Spectral fits of EPIC-pn and EPIC-MOS1+2 data of Obs.~0. When only a continuum and blackbody model is included, the model poorly fits the data (top). We added an absorber, obtaining a significant improvement in the fit quality (middle). Our best fit model is obtained by adding a second absorber to the model (bottom).}
\label{fig:model}
\end{figure}

\begin{figure*}
\centering
\includegraphics[scale=0.35]{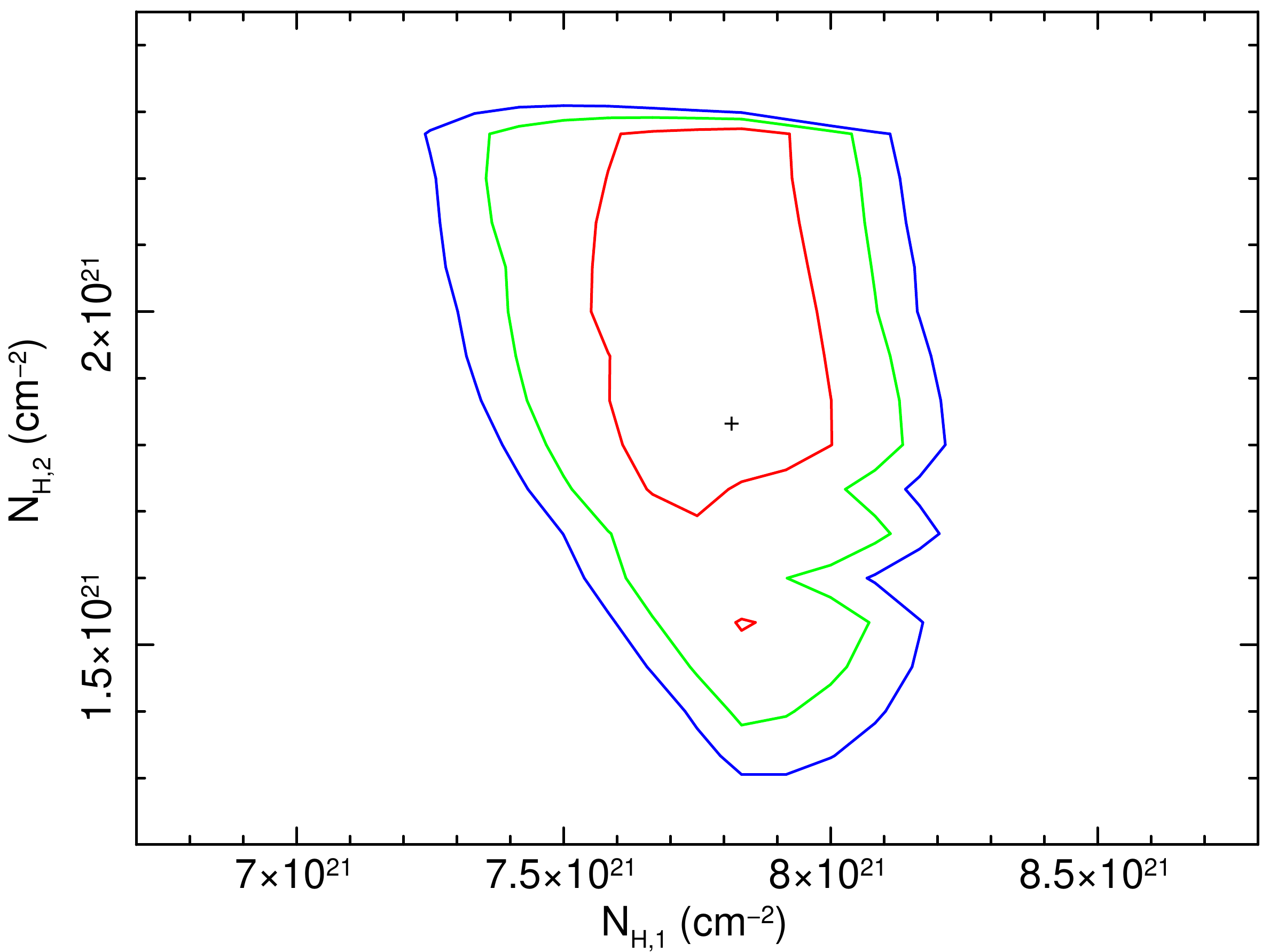}\includegraphics[scale=0.35]{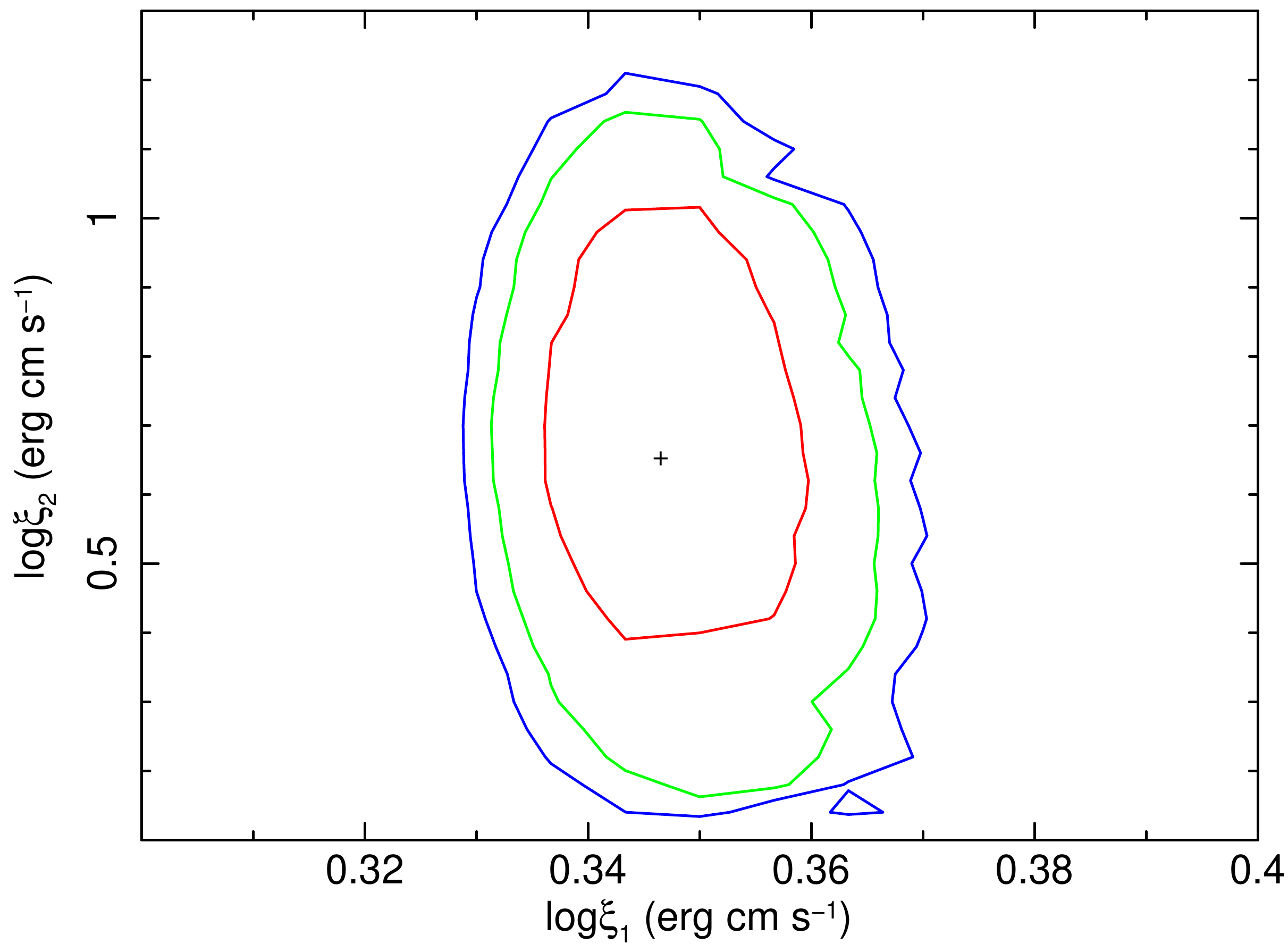}
\caption{Contour plots of column density (left) and ionization parameter (right) of Absorber~1 versus Abs.~2 for Obs.~0. Red, green, and blue lines represent $68\%$, $90\%,$ and $95\%$ confidence levels, respectively. The black cross represents the best-fit values.}
\label{fig:contourabs}
\end{figure*}

\begin{figure}
\centering
\includegraphics[width=\columnwidth]{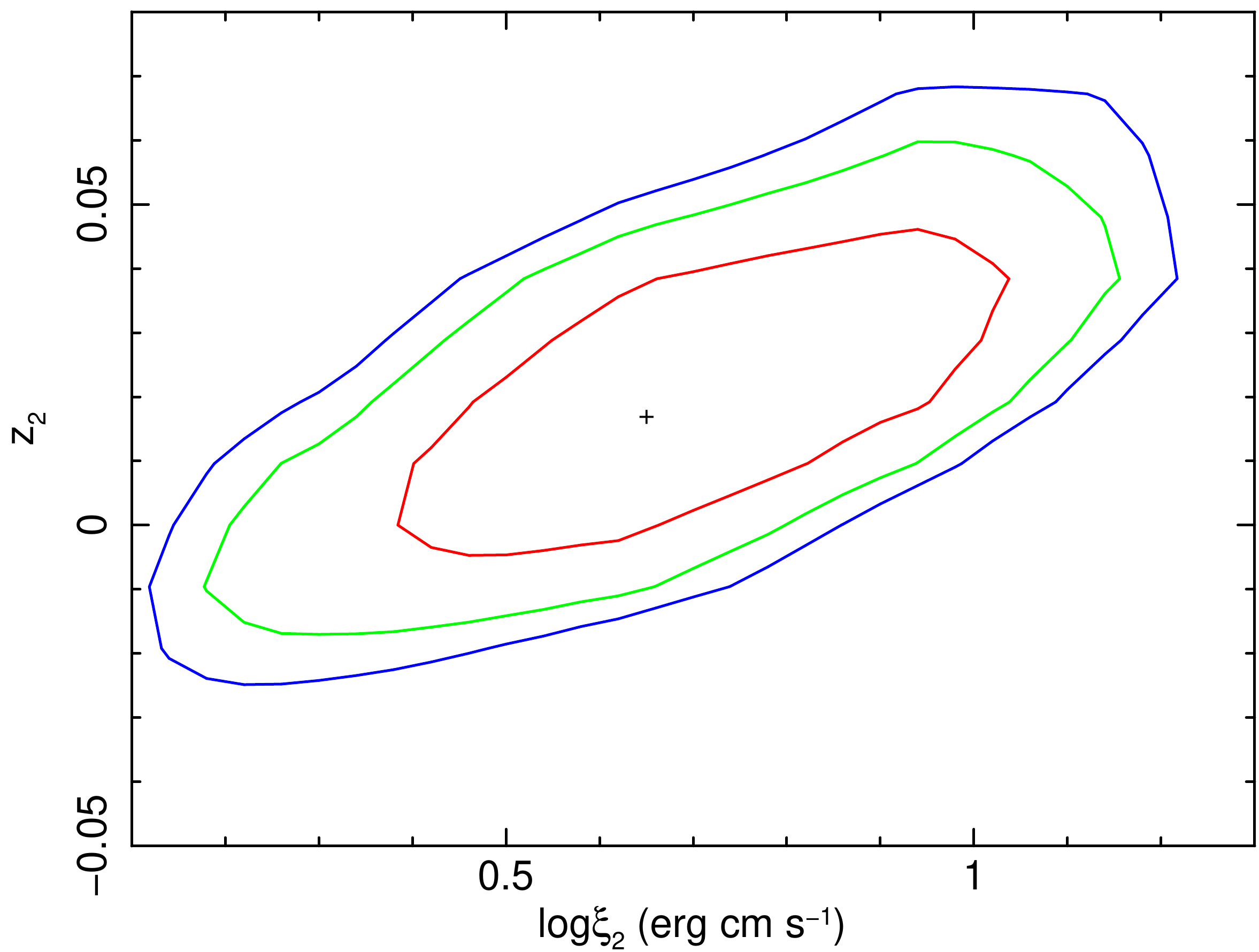}
\caption{Contour plot of the observed redshift of Abs.~2, i.e. $z_{o,2}$ , versus the ionization $\log\xi_2$ for Obs.~0. Red, green, and blue lines represent $68\%$, $90\%,$ and $95\%$ confidence levels, respectively. The black cross represents the best-fit values.}
\label{fig:contourzxi}
\end{figure}

The X-ray spectral analysis was carried out using the software {\footnotesize XSPEC} \citep{arnaud96} included in the High-Energy Astrophysics SOFTware ({\footnotesize HEASOFT}) v6.18 software package. We fitted separately EPIC-pn and EPIC MOS1+2 spectra for each XMM-Newton observation, using a power law model with Galactic absorption, a soft excess modeled by a blackbody spectrum, and two ionized absorbers (named Absorber (Abs)~1 and Abs.~2), following indications from previous analyses of these spectra \citep{ashton04,piconcelli05}.\\
\indent The best fit values of EPIC-pn and MOS spectra agree within a $90\%$ confidence level (see Figs.~\ref{fig:abs2} and~\ref{fig:abs3}), therefore the results were recomputed fitting EPIC-pn and MOS spectra together, to obtain higher significance results. For the ASCA observation, we fitted together the spectra from all cameras of the telescope. Best-fit parameters of each spectra and the improvements in term of $\Delta\chi^2$ for each fit component are listed in Tables~\ref{tab:abs1}-\ref{tab:photonkT}, while spectra and best-fit curves are shown in Figs.~\ref{fig:obs00} to \ref{fig:obs11}. In Fig.~\ref{fig:model}, we show the spectra and ratio of the continuum, continuum+Abs.~1, and continuum+Abs.~1+Abs.~2 models for Obs.~0.\\
\indent We rule out instrumental artifacts or random fluctuations for the detection of Abs.~2 for the following reasons. First, the main spectral features are detected in the observed energy range $E=0.8-1.5$ keV, where no sharp edges are present in the effective area of both pn and MOS instruments. Second, if due to an instrumental artifact, the spectral features would have been detected at the same energy and with the same intensity in many other sources. Third, in case of instrumental artifact, we would not observe the same features in both pn and MOS within a $90\%$ confidence level for all the observations.\\
\indent For three XMM-Newton observations, Obs.~1+2+3, Obs.~6, and Obs.~8+9, we also need a third absorber (Abs.~3) in the Fe K$\alpha$ band, associated with a UFO. In Table~\ref{tab:abs3} we also report the best fit results for a third absorber in the ASCA observation, although it is found with significance below a $2\sigma$ confidence level threshold, in order to compare with the claim of \cite{george97}. We note that the detection probability of all the UFO reported in Table~\ref{tab:abs3} is higher than $99\%$, even when considering a blind line search. In fact, extensive Monte Carlo simulations show that the null probability of a spectral feature in the $E=7-10$ keV band, with $\Delta\text{dof}=3$ (degrees of freedom), is $\sim1\%$ for a $\Delta\chi^2\gtrsim11$ \citep{tombesi10a}.\\
\indent We modeled the ionized absorbers by computing detailed grids with the photoionization code {\footnotesize XSTAR} \citep{kallman01}, which considers absorption lines and edges for every element with $Z\leq30$. For Abs.~1 and 2, we calculated an {\footnotesize XSTAR} table with a spectral energy distribution in the $E=10^{-1} - 10^{6}$ eV energy band described by a photon index of $\Gamma=2$, with cutoff energy beyond $E_c>100$ keV, following \citet{haardt91}. We considered standard solar abundances from \citet{asplund09} and turbulent velocity of $100$ km s$^{-1}$ \citep[e.g.,][]{laha14}. This value is well within the energy resolution provided by the EPIC-pn and MOS instruments, and testing different values did not provide statistically different results. Since Abs.~3 is associated with UFOs, we adopted a nearly identical model, with the only difference being a larger turbulent velocity, $v\sim10^3$ km s$^{-1}$.\\
\indent In the {\footnotesize XSTAR} grids, the free parameters are the absorber column density, $N_\text{H}$, its observed redshift, $z_o$, and the ionization parameter, $\xi=L_\text{ion}/r^2n,$ where $L_\text{ion}$ is the ionizing luminosity between $13.6$ eV and $13.6$ keV, computed by using the {\tt luminosity} task in {\footnotesize XSPEC} on the unabsorbed best fit spectral model. The relation between the observed redshift $z_o$, the cosmological redshift $z_c$, and the Doppler shift $z_a$ of the absorber with respect to the source rest frame is $$1+z_o=(1+z_a)(1+z_c).$$ The velocity of the outflow is related to $z_a$ by the relation $$1+z_a=\sqrt{\frac{1-v/c}{1+v/c}}.$$ In case of outflowing material, $z_a$ is a blueshift, and we conventionally adopt $v/c>0$. 

\subsection{Results}
\label{sec:results}

\begin{figure*}
\centering
\includegraphics[scale=0.7]{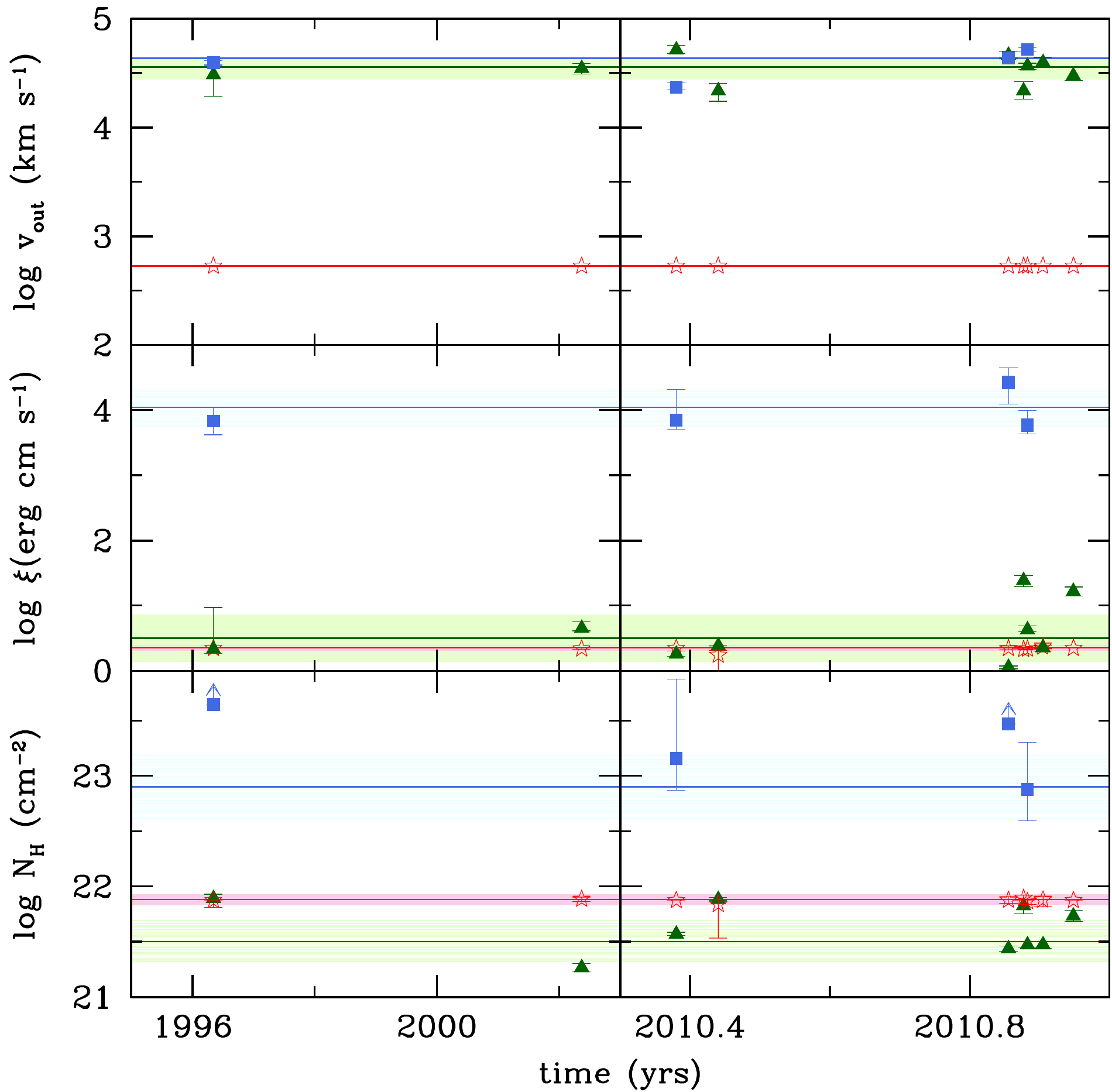}
\caption{Time dependence of the outflow velocity $v_\text{out}$ (top), ionization parameter $\xi$ (middle), and column density $N_\text{H}$ (bottom). For each panel we show the values of Abs.~1 (red stars), Abs.~2 (green triangles), and Abs.~3 (blue squares). Lower limits on $N_\text{H}$ were marked with an arrow. Horizontal lines with shaded bands represent the median values and the corresponding median absolute deviations.}
\label{fig:timeseries}
\end{figure*}

\begin{figure}
\centering
\includegraphics[width=\columnwidth]{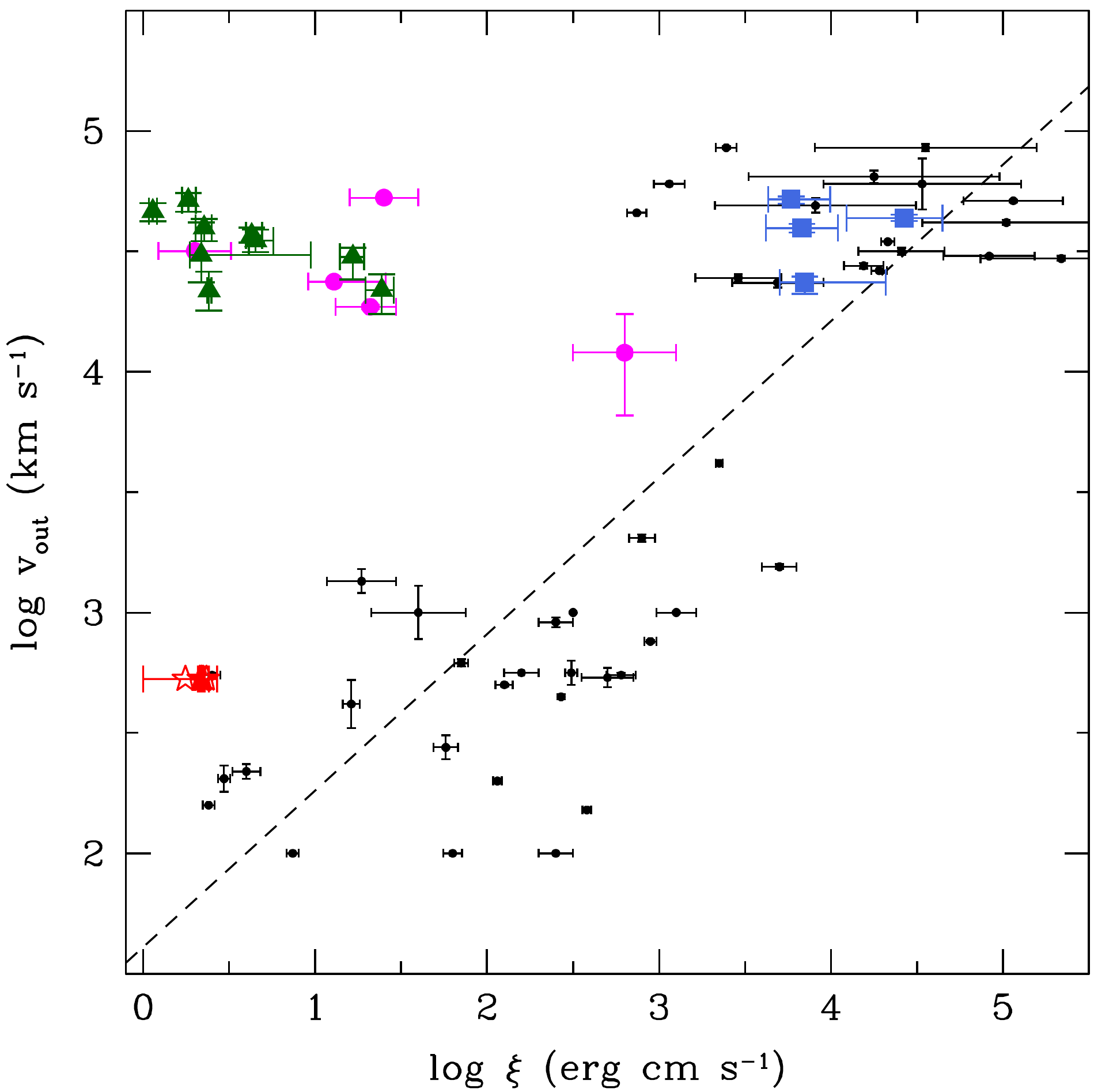}
\caption{Velocity vs. ionization parameter plot for warm absorber (red stars), entrained ultra-fast outflow (green triangles), and ultra-fast outflow (blue squares). As expected, the UFO and the WA seem to follow a linear correlation. The smaller black points and the dashed line represent the points and the linear fit of \citet{tombesi13}. The velocity of the WA constant to the value computed by \citet{mathur98} is $v\sim530$ km s$^{-1}$. The E-UFO does not follow such a correlation. The magenta points represent other soft X-ray UFOs reported in the literature (see Sect.~\ref{sec:intro} for details).}
\label{fig:voutvxi}
\end{figure}

\begin{figure}[h!]
\centering
\includegraphics[width=\columnwidth]{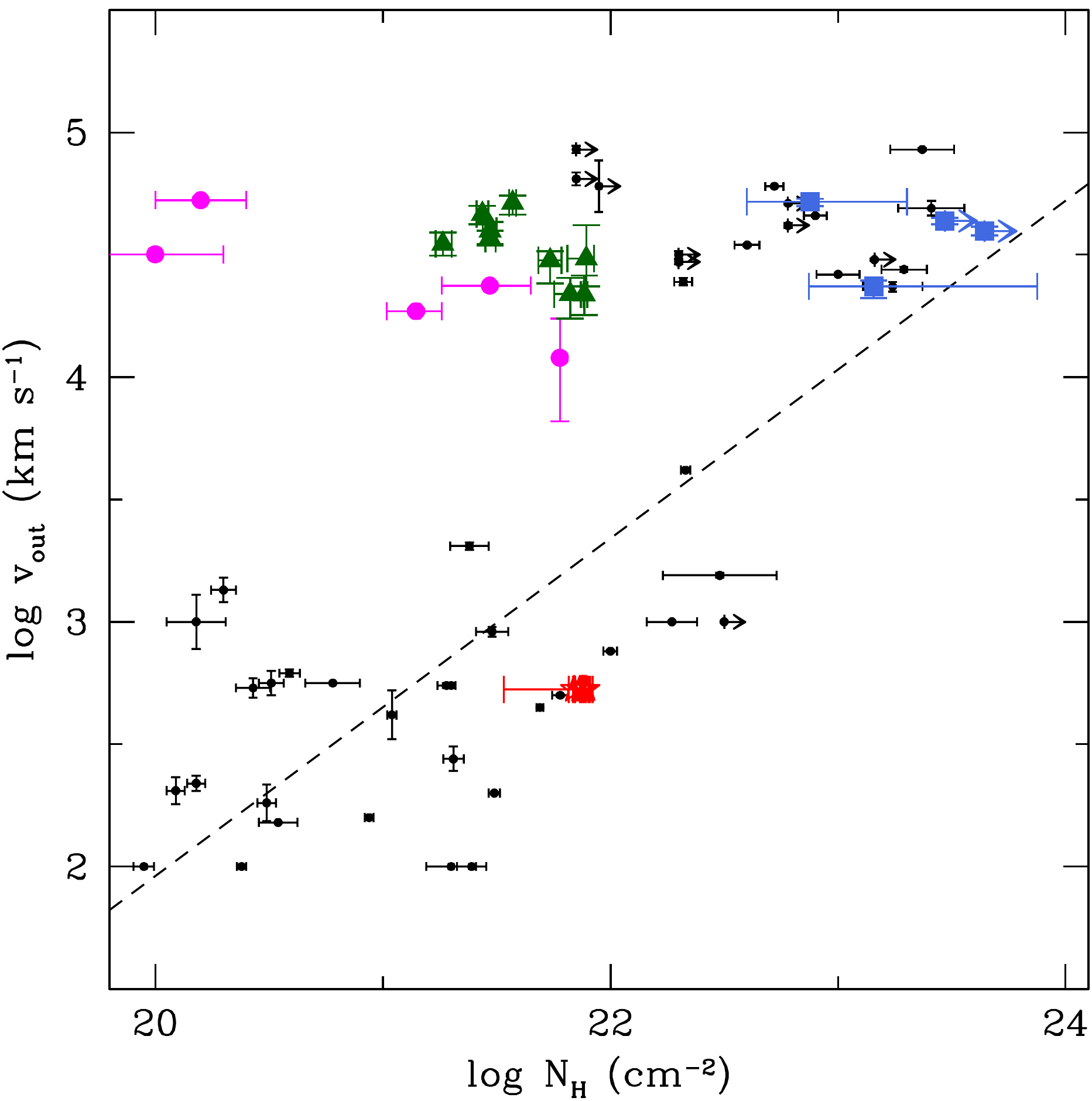}
\caption{Velocity vs. column density plot for WA (red stars), E-UFO (green triangles), and UFO (blue squares). As in the case of Fig.~\ref{fig:voutvxi}, the warm absorber and the ultra-fast outflow are in agreement with \citet{tombesi13} (smaller black dots and dashed line). Arrows represent lower limits on $N_\text{H}$. Magenta points are soft X-ray UFOs in the literature. Again, the velocity of the warm absorber is assumed to be $v\sim530$ km s$^{-1}$. Even though the green triangles do not follow the linear fit of \citet{tombesi13}, there seems to be some continuity between the column density of this absorber and the `regular' ultra-fast outflows.}
\label{fig:voutvnh}
\end{figure}

\begin{figure}[h!]
\centering
\includegraphics[width=\columnwidth]{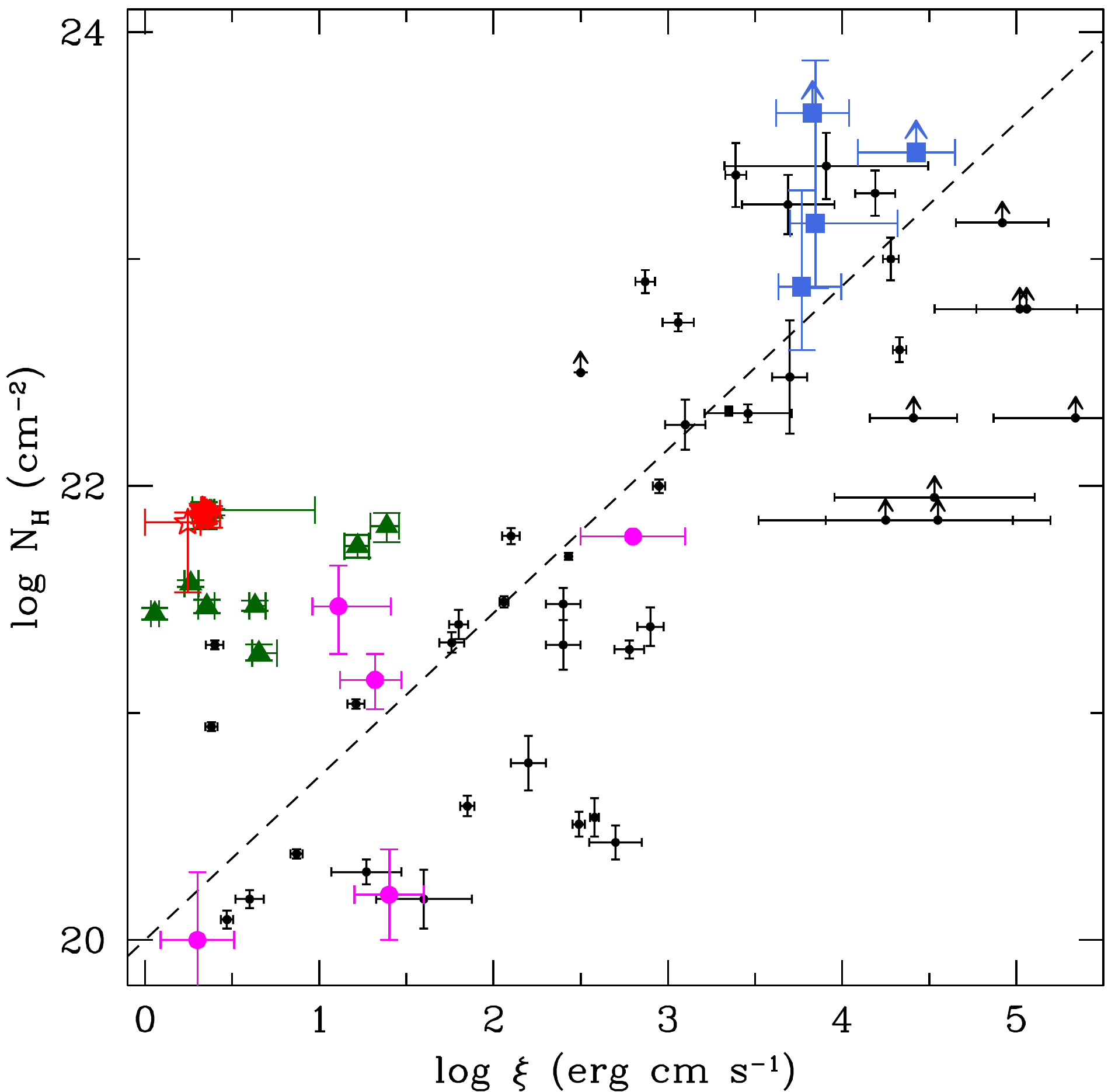}
\caption{Column density vs. ionization parameter plot for WA (red stars), E-UFO (green triangles), and UFO (blue squares). Arrows represent lower limits on the column density, with the black dots and dashed line representing the linear fit from \citet{tombesi13}. Magenta points are soft X-ray UFOs from the literature. This plot does not show any difference from \citet{tombesi13}, since the E-UFO only differs from the WA in its velocity.}
\label{fig:nhvxi}
\end{figure}

Absorber~1 shows a very low variability (see Table~\ref{tab:abs1}) in both ionization parameter $\log (\xi_1/\text{erg cm s}^{-1})=0.35\pm0.04$ and column density $\log (N_{H,\text{1}}/\text{cm}^{-2})=21.88\pm0.05$. The redshift of the absorber, mainly due to the shift of the Fe M-shell unresolved transition array (UTA), is comparable with the cosmological one of the source ($z_o=z_c$) within the EPIC energy resolution. In fact, when the absorber redshift is left free, the difference between the source redshift and the observed one is on the order of $\Delta z\sim10^{-3}$, far below the energy resolution of the instrument. Therefore, its velocity is assumed to be fixed at $v_{1,\text{out}}\sim530$ km s$^{-1}$, as shown by its corresponding absorption phase in the UV band \citep{mathur98}. These values of column density, velocity, and ionization are typical of a WA \citep{blustin05}.\\
\indent Absorber~2 is more variable, as shown in Table~\ref{tab:abs2}, with median values for the ionization parameter and column density given by $\log(\xi_2/\text{erg cm s}^{-1})=0.50\pm0.36$ and $\log (N_{\text{H},2}/\text{cm}^{-2})=21.5\pm0.2$, respectively. We verified that the values of the parameters of Abs.~2 are independent from the ones of Abs.~1, ruling out any systematically induced correlation. In fact, even if the median values of the two absorbers are of the same order, the values of the individual observations are not correlated. Indeed, the correlation coefficient between non-fixed values of $N_\text{H,1}$ and the corresponding $N_\text{H,2}$ is $r=0.14$, while the one between $\log\xi_1$ and $\log\xi_2$ is $r=0.17$, showing that their variability is not related. Moreover, contour plots show no significant correlation between the column density and ionization parameter of Abs.~1 and Abs.~2 (see Fig.~\ref{fig:contourabs}). The velocity of this absorber is measured with respect to the centroid energy of UTA Fe M and oxygen lines for the observations with lower ionization parameter, and the addition of Fe L lines for the observations with higher ionization parameter, such as Obs.~7 and 11 (see Table~\ref{tab:abs2}). Given the low resolution of the EPIC cameras, however, only Fe UTA are likely to contribute to the redshift calculation. As shown in Fig.~\ref{fig:contourzxi}, contour plots of the observed redshift $z_{o,2}$ and $\log\xi_2$ are well constrained. However, the observed redshift is below the systemic value $z_c=0.144$ even at $95\%$ confidence level. The median value of the velocity of this absorber is $v_{2,\text{out}}=(0.120\pm0.029)c.$ Given the high velocity and low values of the ionization and column density, this absorber shows intermediate parameters between a WA and a UFO. In fact, outflows with $v\sim0.1-0.4c$ are usually coupled with ionization parameters in the range $\log\xi/\text{erg~cm~s}^{-1}\sim3-6$ \citep{tombesi13}, detectable by observing Fe {\footnotesize XXV} and {\footnotesize XXVI} absorption lines. The ionization parameter of this absorber is instead in the range $\log\xi_2/\text{erg~cm~s}^{-1}\sim0-1.5$, much lower than the usual UFO and comparable with the range of ionization we find in WAs. \\
\indent As mentioned, three observations show evidence of a third absorber. This absorber is variable (see Table~\ref{tab:abs3}) and the median values of column density, ionization, and velocity are respectively $\log(N_{H,\text{3}}/\text{cm}^{-2})=22.9\pm0.3$, $\log(\xi_3/\text{erg cm s}^{-1})=4.04\pm0.29,$ and $v_{3,\text{out}}=(0.140\pm0.035)c$. These values are consistent with those of a typical UFO \citep[e.g.,][]{tombesi11,king15}. The median values of all the parameters of the three absorbers are summarized in Table~\ref{tab:medianvalues}, while the best-fit results of each absorber in time are shown in Fig.~\ref{fig:timeseries}.\\
\indent While the WA and the UFO agree with the linear trends in \cite{tombesi13} between column density, ionization parameter, and velocity, Abs.~2 does not fit well into such relations (see Fig.~\ref{fig:voutvxi},~\ref{fig:voutvnh}, and~\ref{fig:nhvxi}) and it lies in a different region of the plots. It is straightforward to note that the velocity of Abs.~2 and that of the UFO are consistent within their dispersion, while $\xi$ and $N_H$ are consistent with those of the WA. This is a strong indication that this absorber is likely to be an intermediate phase between these two, possibly related to the interstellar medium being entrained by the UFO (see Sect.~\ref{sec:winden}), and hence we can name this absorber the entrained ultra-fast outflow (E-UFO).

\section{Distances of the absorbers}
\label{sec:dist}

{\renewcommand{\arraystretch}{1.2}
\begin{table}
\centering
\caption{Median values and median absolute deviation (MAD) for each parameter of three absorbers using the best-fit values of the XMM-Newton spectra.}
\label{tab:medianvalues}
\begin{tabular}{lccc}
\hline
Parameter & Median & Units\\
\hline
$\log N_{\text{H},1}$ & $21.88\pm0.05$ & cm$^{-2}$\\
$\log\xi_1$ & $0.35\pm0.04$ & erg cm s$^{-1}$\\
$v_{1,\text{out}}{*}$ & $\sim530$ & km s$^{-1}$\\
$\log N_{\text{H},2}$ & $21.5\pm0.2$ & cm$^{-2}$\\
$\log\xi_2$ & $0.50\pm0.36$ & erg cm s$^{-1}$\\
$v_{2,\text{out}}$ & $0.120\pm0.029$ & $c$\\
$\log N_{\text{H},3}{**}$ & $22.9\pm0.3$ & cm$^{-2}$\\
$\log\xi_3$ & $4.04\pm0.29$ &  erg cm s$^{-1}$\\
$v_{3,\text{out}}$ & $0.145\pm0.035$ & $c$\\
$\log L_\text{ion}$ & $45.4\pm0.3$ & erg s$^{-1}$\\
\hline
\end{tabular}
\tablefoot{We also report the median ionizing luminosity $L_\text{ion}$. They were computed putting together lower and upper limits of each parameter of the XMM-Newton observations. Units are reported; in the case of logarithmic parameters, the units are to be understood as related to the argument. ${*}$ The value of $v_{1,\text{out}}$ was set to $530$ km s$^{-1}$, following \citet{mathur98}. ${**}$ We computed median and MAD for the lower values of $N_{\text{H},3}$, since we only have lower limits for two observations (see Table~\ref{tab:abs3}).}
\end{table}
}

Following \citet{tombesi13} and \citet{crenshaw12} we can estimate the maximum distance of the absorber from the black hole, by assuming that the shell size is smaller than its distance from the center, meaning $N_H=nR<nr$, with $R$ size of the shell and $r$ distance from the SMBH, so that $n=N_H/r_\text{max}$. The determination of a well-defined ionization parameter indicates that we are observing a shell of gas with a thickness much smaller than the average distance of the absorber to the source, otherwise we would have detected a significantly larger gradient in the ionization parameter. Therefore we have

\begin{equation}
r_\text{max}=\frac{L_\text{ion}}{N_\text{H}\xi},
\label{eq:rmax}
\end{equation}

\noindent where $L_\text{ion}$ is the unabsorbed ionizing luminosity emitted by the source, for which we used the median value reported in Table~\ref{tab:medianvalues}.\\
\indent We are also able to compute the minimum distance from the black hole, by computing the distance at which the observed velocity is equal to the escape velocity,

\begin{equation}
r_\text{min}=\frac{2GM_\text{BH}}{v_\text{out}^2}.
\label{eq:rmin}
\end{equation}

\noindent For the warm absorber, we obtain $r_\text{min,WA}\sim10^5r_s\simeq18$ pc and $r_\text{max,WA}\sim10^9r_s\simeq5\times10^5$ pc, where $r_s=2GM_\text{BH}/c$ is the Schwarzschild radius. Since no constraints can be put on these values, we have to assume that the WA is located beyond $r_\text{min,WA}$. However, the external radius $r_\text{max,WA}$ is too extreme and therefore we do not report it on Table~\ref{tab:energetics}. We assume instead a typical value $r_\text{max,WA}\sim3$ kpc \citep{digesu13}.\\
\indent For the UFO, we obtain $r_\text{min,UFO}\sim57r_s\simeq3\times10^{-3}$ pc and $r_\text{max,UFO}\sim10^4r_s\simeq0.5$ pc. Numerical simulations \citep[e.g.,][]{fukumura15,nomura16,sadowski17} show that the UFO launching region is confined within $r\lesssim100r_s$ and therefore we conservatively assume $r_\text{min,UFO}$ as the typical value of the distance of the UFO from the central SMBH.\\
\indent Having a UFO-like velocity, the minimum distance of the E-UFO from the central SMBH is much smaller than that of the WA, which is $r_\text{min,E-UFO}\sim70r_s$. However, this value that is consistent with the distance of the UFO is not applicable, since a clumpy wind co-spatial with the UFO at $r<100r_s$ \citep[e.g.,][]{pounds16} is excluded on the basis of photoionization considerations. In fact, the absorber would require a Compton-thick column density in order to be shielded from the intense radiation field and at the same time preserve its low ionization parameter. Moreover, the consistent values of both $N_\text{H}$ and $\xi$ between the WA and the E-UFO strongly suggest that the two absorbers share the same material, hence they are at comparable distances from the SMBH. Therefore, the E-UFO is most likely located in a shell with $r_\text{min,E-UFO}\gtrsim18$ pc, in agreement with the location of the WA.\\
\indent An alternative method to compute the distance of the absorber from the SMBH is to use the median absolute deviations (MAD) of $N_\text{H}$ and $v_\text{out}$ to estimate a typical value of the density of the absorber. Since a significant degree of clumpiness is expected for the ambient medium \citep[e.g.,][]{gaspari17}, it is reasonable to assume that the variations of the typical shell radius $R$ are due to the velocity dispersion $\Delta v_\text{out}$. However, this is due to the fact that the shell's apparent thickness depends on the peculiar clump we are observing, and it is not due to an intrinsic variability of the velocity and the physical properties of the gas in the clump.\\
\indent To derive an estimate of the shell's apparent thickness we consider $N_\text{H}\simeq nR$, and we can obtain through partial derivation 
\begin{equation}
\frac{\Delta N_\text{H}}{\Delta t}=n\frac{\Delta R}{\Delta t}+R\frac{\Delta n}{\Delta t}.
\end{equation}
\indent We assume that in the time span of the XMM-Newton observations, $\Delta t\simeq8.5$ yrs, the average density of the shell does not significantly vary, therefore $\Delta n/\Delta t\simeq0$, and we assume that $\Delta R/\Delta t$ corresponds to the MAD of the velocity of the absorber, $\Delta v_\text{out}$. Given the variability of $N_\text{H}$ and $v_\text{out}$, we are able to compute a typical value of the shell density and consequently the distance of the absorber from the black hole, using the following equations:

\begin{equation}
\left < n \right >=\frac{\Delta N_\text{H}}{\Delta t \Delta v_\text{out}}
\label{eq:density}
,\end{equation}

\begin{equation}
r_\text{var}=\sqrt{\frac{L_\text{ion}}{\left < n \right >\xi}}.
\label{eq:rvar}
\end{equation}

\noindent The distance of the E-UFO from the SMBH is therefore $r_\text{var,E-UFO}\sim2\times10^6r_s\simeq109$ pc, which is consistent with being at least partially co-spatial with the WA at $r\gtrsim18$ pc.\\
\indent However, these distance estimates are representative of the location of the three absorbers, but it is not physically required to define a division between the regions, given that there is unlikely to be a sharp physical discontinuity among them. All the values of the distance from the central black hole and density are summarized in Table~\ref{tab:energetics}.

\section{Outflow energetics}
\label{sec:winden}

{\renewcommand{\arraystretch}{1.5}
\begin{table*}
\centering
\caption{Distance, density, mass outflow rate, momentum rate and kinetic power for each absorber.}
\label{tab:energetics}
\begin{tabular}{lcccccc}
\hline
 & $r/r_\text{s}$ & $r$ (pc) & $n$ (cm$^{-3}$) & $\dot{M}_\text{out}$ ($M_\odot$/yr) & $\dot{P}_\text{out}/\dot{P}_\text{rad}$ & $\dot{E}_\text{K}/L_\text{bol}$\\
\hline
{\bf Warm absorber}\\
$r_\text{min,WA}$ & $3.2\times10^5$ & $18$ & $3.6\times10^5$ & $2.63\,C_v$ & $0.05\,C_v$ & $4.2\times10^{-5}\,C_v$\\
$r_\text{max,WA}$ & $5.2\times10^7$ & $3\times10^3$ & $13$ & $430\,C_v$ & $8\,C_v$ & $7\times10^{-3}\,C_v$\\
{\bf Entrained UFO}\\
$r_\text{var,E-UFO}$ & $1.7\times10^6$ & $110$ & $7.5\times10^3$ & $400\,C_v$ & $495\,C_v$ & $29\,C_v$\\
{\bf Ultra-fast outflow}\\
$r_\text{min,UFO}$ & $57$ & $3.2\times10^{-3}$ & $2.3\times10^9$ & $0.41$ & $0.62$ & $0.04$\\
\hline
\end{tabular}
\tablefoot{The distance from the black hole $r$ is given in both units of Schwarzschild radius $r_s=2GM_\text{BH}/c^2$ and parsec. We listed the estimated density $n$ at such a distance, mass outflow rate in units of solar masses per year, momentum rate in units of the radiation momentum rate $\dot{P}_\text{rad}=L_\text{bol}/c$, and kinetic power in units of $L_\text{bol}$. The symbol $C_v$ represents the filling factor of the region, assumed unitary for the region containing the UFO. We use $M_\text{BH}\sim5.9\times10^8M_\odot$ and $L_\text{bol}\sim5.5\times10^{45}$ erg s$^{-1}$ from \citet{shen11}. For UFO and E-UFO we only reported the most likely value for the distance of the absorber, while for the WA it is not possible to constrain the region, therefore we report the values computed using $r_\text{min}$, and consider a typical value for the upper limit of $r_\text{max,WA}\sim3$ kpc \citep{digesu13}.}
\end{table*}
}

The mass outflow rate can be estimated following the equation in  \citet{crenshaw12}, 

\begin{equation}
\dot{M}_\text{out}=4\pi\mu m_\text{p}C_vC_fN_\text{H}v_\text{out}r,
\label{eq:massoutflowrate}
\end{equation}

\noindent where $m_\text{p}$ is the proton mass, while $\mu\equiv n_\text{H}/n_e\simeq1/1.4$ for solar abundances. The symbol $C_f$ represents the covering factor, which we assume to be $C_f\simeq0.5$ \citep[e.g.,][]{tombesi10a}, and $C_v$ is the filling factor of the region, which we assume to be unitary for the UFO, according to previous results \citep[e.g., ][]{nardini15,tombesi15}, and that we keep as parametric for both the E-UFO and WA. We also assume that E-UFO and WA also share the same clumpiness.\\
\indent The momentum rate of the outflow is given by
\begin{equation}
\dot{P}_\text{out}=\dot{M}_\text{out}v_\text{out}.
\end{equation}
\indent Finally, we compute the kinetic power
\begin{equation}
\dot{E}_\text{K}=\frac{1}{2}\dot{M}_\text{out}v_\text{out}^2.
\label{eq:kineticpower}
\end{equation}
\indent We compute Eqs.~\ref{eq:massoutflowrate}-\ref{eq:kineticpower} for each of the three absorbing complexes, using the median values of $N_\text{H}$ and $v_\text{out}$ (see Table~\ref{tab:medianvalues}) and the corresponding distance as computed in Sect.~\ref{sec:dist}. The complete set of $r$, $\dot{M}_\text{out}$, $\dot{P}_\text{out}$ , and $\dot{E}_\text{K}$ values is listed in Table~\ref{tab:energetics}, with $\dot{P}_\text{out}$ in units of the momentum rate of the AGN, $\dot{P}_\text{rad}=L_\text{bol}/c$, with $L_\text{bol}=5.5\times10^{45}$ erg s$^{-1}$ \citep{shen11}, and $\dot{E}_\text{K}$ in units of $L_\text{bol}$.
\indent The median velocities of E-UFO and UFO are perfectly comparable within the errors. This suggests that these two absorbers are dynamically connected, which means that the two ionized absorbers have interacted. If we assume an energy-conserving interaction between the UFO and the E-UFO, the filling factor of the region containing the entrained outflow must be $C_v\simeq1.4\times10^{-3}$. However, we also obtain a very similar value assuming a momentum-conserving interaction between the UFO and the E-UFO. The density of the E-UFO region, as computed in Eq.~\ref{eq:density}, is $\left < n_2 \right >\simeq7.5\times10^3$ cm$^{-3}$. If we assume for the UFO a distance from the black hole $r_3\geq r_\text{min}$ , we obtain $n_3\leq L_\text{ion}/\xi r_\text{min}^2\simeq2.3\times10^9$ cm$^{-3}$. Assuming a mass-conserving spherical shell, the density scales as $n\propto r^{-2}$, and the predicted value of the density at $r_\text{var}\simeq110$ pc would be $n_2\lesssim120$ cm$^{-3}$, around a factor of $\sim100$ lower than the one obtained with variability arguments.  This means that the average density computed with Eq.~\ref{eq:density} is dominated by the density of the clumps in the E-UFO region. Moreover, if we assume that $N_\text{H}\propto n$, since $N_{\text{H},3}\simeq5\times10^{23}$ cm$^{-2}$, the UFO column density at $r_\text{var}$ would be $N_\text{H}\sim10^{16}$ cm$^{-2}$, much lower than the observed $N_{\text{H},2}\simeq3.1\times10^{21}$ cm$^{-2}$, which supports our requirement for clumpiness. This clumpiness is supported by detailed models of AGN feedback driven by outflows \citep[][]{zubovas14,costa14,gaspari17}, in which Rayleigh-Taylor and Kelvin-Helmholtz instabilities, together with radiative cooling, cause the shell of shocked gas to fragment into clumpy warm and cold clouds that fill the nuclear and kiloparsec region. The clumpy environment and ensemble of cold or warm clouds is more in line with with the CCA scenario \citep{gaspari13} instead of spherical hot-mode accretion \citep{bondi52}. The typical core filling factor, obtained in simulations \citep{gaspari17b}, is $\sim0.1-1\%$, and supports our result. The mass outflow rate of the E-UFO is $\dot{M}_{\text{out},2}\simeq0.55 M_\odot/\text{yr}$, which is $\sim35\%$ larger than the UFO value $\dot{M}_{\text{out},1}\simeq0.41 M_\odot/\text{yr}$, possibly meaning that the mass is growing by sweeping-up the material that is being entrained by the UFO.  \\
\indent Given that WA and E-UFO share the same gas properties (i.e., ionization and column density), we also assume that the two regions share the same filling factor, in which case we obtain $\dot{P}_{\text{out},1}\ll\dot{P}_{\text{out},3}$ and $\dot{E}_{\text{K},1}\ll\dot{E}_{\text{K},3}$, which is compatible with the fact that the two absorbers are not dynamically related, suggesting that the WA represents an unperturbed region that has not yet been significantly affected by the UFO.

\section{Summary and discussion}
\label{sec:discussion}

\begin{figure*}
\centering
\includegraphics[scale=0.35]{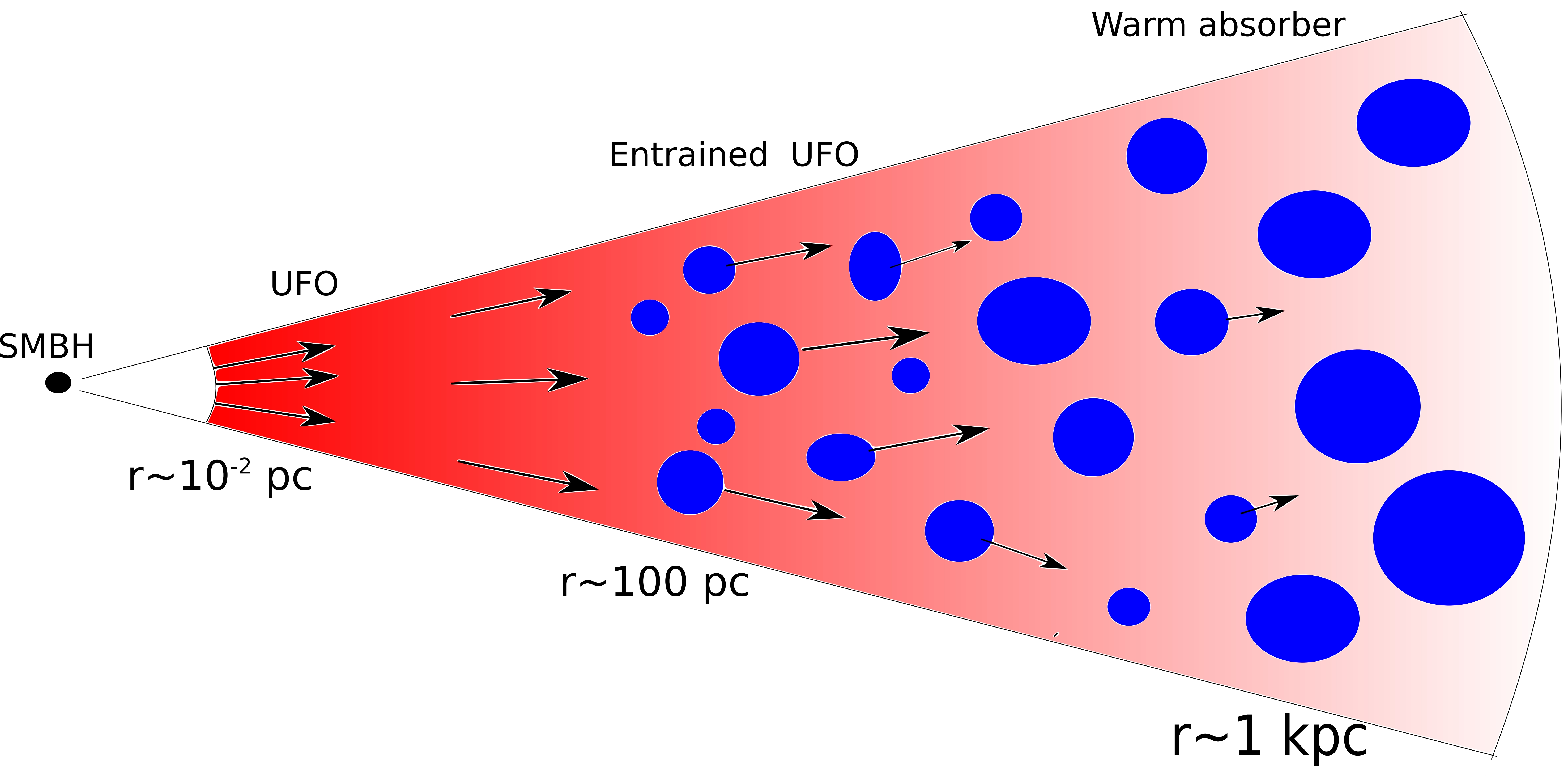}
\caption{Diagram of the X-ray observations of PG 1114+445. A UFO is present in the inner part of the AGN surroundings, with decreasing density, scaling as $r^{-2}$. At larger distances from the SMBH, the UFO interacts with the closest clumpy ambient gas at $r\simeq100$ pc, entraining it via Rayleigh-Taylor and Kelvin-Helmholtz instabilities. This gas is pushed at velocities comparable with that of the UFO, retaining its ionization state and column density. The farther ambient gas remains unaffected by the UFO and therefore moves at a significantly lower line-of-sight velocity. The figure is not to scale.}
\label{fig:windscheme}
\end{figure*}

We summarize our findings in graphical form by presenting the multiphase absorber of PG 1114+445 in the diagram shown in Fig. \ref{fig:windscheme}. Together with the analysis described in this work, such a scenario is also well supported by theoretical models of AGN outflow evolution \citep[e.g.,][]{king03,king05,zubovas12}, which predict an inner UFO driving into the surrounding medium, sweeping-up the gas outwards. In the aftermath of the shock, four regions are consequently formed: (i) the innermost region, containing the unshocked UFO, (ii) the shocked inner wind, (iii) the shock-induced swept-up interstellar gas, and (iv) the unaffected ambient medium.\\
\indent A UFO with high velocity ($v_{\text{out},3}\simeq0.145c$), ionization parameter, and column density \citep[e.g., ][]{tombesi11} is detected in three XMM-Newton observations. This absorber is located very close to the source, at $r\sim3\times10^{-3}\text{ pc}\simeq60r_\text{s}$ (see Table~\ref{tab:energetics}), and can be associated with the inner wind, in agreement with detailed general-relativistic radiative magnetohydrodynamic (GR-rMHD) simulations \citep[e.g.,][]{sadowski17}. On the contrary, given its expected low particle density (see Sect.~\ref{sec:winden}), the shocked inner wind is likely to be undetectable with the present dataset, since an expected column density of $N_\text{H}\sim10^{16}$ cm$^{-2}$ is below the sensitivity limit of current spectrometers.\\
\indent The low values of $v$, $\log\xi$, and $N_\text{H}$ for the WA are compatible with usual absorbers of this kind \citep[e.g., ][]{blustin05}. Its low variability (see Fig.~\ref{fig:timeseries}) suggests that the WA is not dynamically connected to the inner UFO. In fact, WAs are believed to be relics of the original optically thick gas surrounding the black hole before any AGN activity started. This gas may have been driven towards larger distances by the initial radiation pressure of the quasar, until it became optically thin \citep{king14}. Alternatively, the ambient clouds are drifting in the macro-scale turbulent velocity field \citep{gaspari17}, with velocities of a few hundred km s$^{-1}$. Supported by the constancy over more than a decade of observations, the WA is most likely part of the ambient medium, not yet significantly affected by the UFO.\\
\indent A comparison between the $v-\xi$, $v-N_\text{H}$ and $\xi-N_\text{H}$ relations for the three absorbers (see  Figs.~\ref{fig:voutvxi},~\ref{fig:voutvnh}, and~\ref{fig:nhvxi}) shows that the WA and the UFO are well around the expected values \citep{tombesi13}, whereas, on the other hand, the E-UFO has the typical velocity of a UFO, but ionization and column density values comparable with the WA.\\
\indent We interpret these three main absorbers phases as the time evolution of outflows that were expelled at different epochs from the SMBH accretion disk and are observed as the material is continuing to travel outwards. These considerations are valid under the plausible assumption that the mass accretion rate of the SMBH, and consequently the power of the UFO, did not significantly change during a timescale of $t\sim r/v\sim100\text{ pc}/0.145c\sim2,000$ years, which is negligible compared to the minimum time required for the SMBH to double its mass, equal to the Salpeter time $t_S\sim5\times10^7$ yrs. Moreover, we note that in low-mass galaxy haloes, the AGN accretion duty cycle is almost quasi-continuous, thus allowing for the observation of multiple generations of AGN feedback events \citep{gaspari17}. Finally, the interaction between the UFO and the clumpy ambient medium is most likely Rayleigh-Taylor and Kelvin-Helmholtz unstable \citep[e.g.,][]{king10a,zubovas14,costa14}, consistently with the variability of the E-UFO and its low volume filling factor $C_v$.\\
\indent Our detection of three concurrent outflow phases in the same source provides a remarkable corroboration of the self-regulated mechanical AGN feedback scenario \citep[e.g.,][]{king03,king05,king14,gaspari17}. Similar conclusions, however not considering a clumpiness of the ambient medium, have been reached by \cite{sanfrutos18}, for IRAS 17020+4544, in which all three phases have also been detected. The UFO, driven by the accreting SMBH, propagates to meso scales (parsec to kiloparsec), still with significant velocity, where the entrainment and mixing with the turbulent and clumpy ambient medium becomes substantial (E-UFO). The outflow kinetic energy is then likely deposited at macro scales in the form of thermal energy, via bubbles, shocks, and turbulent mixing \citep{gaspari17,lau17}, thus re-heating the galaxy and the group gaseous halo. This self-regulated cycle may happen many times during the AGN's lifetime, giving rise to a symbiotic relation between the SMBH growth and the evolution of the host galaxy \citep{gaspari17b}.\\
\indent The combination of multiepoch observations of nuclear winds and multiwavelength investigations of spectral features tracing multiphase parsec to kiloparsec outflows \citep[e.g.,][]{fiore17} can be considered as the most promising strategy to shed light on the AGN feedback processes over a large range of distances from the central engine. In particular, UV and optical spectra of PG 1114+445 are currently being analyzed in order to link the physical properties of outflows in different spectral bands and will be the subject of upcoming articles.

\begin{acknowledgements}
      We thank Valentina Braito, James Reeves, Cristian Vignali, Jelle Kaastra, Andrew King, Andrew Lobban, and Riccardo Middei for discussions and suggestions. We thank the referee for useful suggestions that improved the quality of this article. RS, FV, and EP acknowledge financial contribution from the agreement ASI-INAF n.2017-14-H.0. FT acknowledges support by the Programma per Giovani Ricercatori - anno 2014 “Rita Levi Montalcini”. MG is supported by the Lyman Spitzer Jr. Fellowship (Princeton University) and by NASA Chandra grants GO7-18121X and GO8-19104X. The results are based on observations obtained with XMM-Newton, an ESA science mission with instruments and contributions directly funded by ESA Member States and NASA. This research has also made use of data and software provided by the High Energy Astrophysics Science Archive Research Center (HEASARC), which is a service of the Astrophysics Science Division at NASA/GSFC and the High Energy Astrophysics Division of the Smithsonian Astrophysical Observatory.
\end{acknowledgements}

% WARNING
%-------------------------------------------------------------------
% Please note that we have included the references to the file aa.dem in
% order to compile it, but we ask you to:
%
% - use BibTeX with the regular commands:
   \bibliographystyle{aa} % style aa.bst
   \bibliography{biblio} % your references Yourfile.bib
%
% - join the .bib files when you upload your source files
%-------------------------------------------------------------------
\onecolumn

\begin{appendix}

\section{Data}
\label{app:data}

{\renewcommand{\arraystretch}{1.5}
\begin{table*}[h!]
\caption{Data considered in this work for PG 1114+445.}
 \label{tab:sample}
        \centering
        \begin{tabular}{ccccccc} % four columns, alignment for each
                \hline
                ID & OBSID & Start date & Instrument &  Exposure time (s) & Counts (0.3-10 keV)\\
                \hline
A & 74072000 & 1996-05-05 23:59:48 & ASCA sis0+1 & 121230 & 6089*\\
& & & ASCA gis2+3 & 136800 & 6138*\\
0 & 0109080801 & 2002-05-14 15:27:00 & EPIC-pn & 32500 & 23462\\
& & & MOS1+2 & 80380 & 17700\\
1 & 0651330101 & 2010-05-19 09:48:59 & EPIC-pn & 20550 & 7166\\
& & & MOS1+2 & 57270 & 6418\\
2 & 0651330201 & 2010-05-21 09:41:13 & EPIC-pn & 9824 & 3924\\
& & & MOS1+2 & 27400 & 3213\\
3 & 0651330301 & 2010-05-23 10:06:23 & EPIC-pn & 5271 & 1669\\
& & & MOS1+2 & 18357 & 1792\\
4 & 0651330401 & 2010-06-10 07:28:14 & EPIC-pn & 10740 & 5341\\
& & & MOS1+2 & 32750 & 5214\\
5 & 0651330501 & 2010-06-14 07:56:46 & EPIC-pn & 6613 & 2956\\
& & & MOS1+2 & 16804 & 2314\\
6 & 0651330601 & 2010-11-08 23:22:49 & EPIC-pn & 18500 & 17359\\
& & & MOS1+2 & 46400 & 12704\\
7 & 0651330701 & 2010-11-16 22:50:52 & EPIC-pn & 16240 & 9732\\
& & & MOS1+2 & 46510 & 9046\\
8 & 0651330801 & 2010-11-18 22:42:54 & EPIC-pn & 20350 & 9707\\
& & & MOS1+2 & 57020 & 8490\\
9 & 0651330901 & 2010-11-20 22:35:32 & EPIC-pn & 21350 & 12244\\
& & & MOS1+2 & 52210 & 9724\\
10 & 0651331001 & 2010-11-26 23:40:17 & EPIC-pn & 17660 & 8704\\
& & & MOS1+2 & 46040 & 7173\\
11 & 0651331101 & 2010-12-12 22:31:31 & EPIC-pn & 13710 & 7439 \\
& & & MOS1+2 & 26650 & 4562\\
                \hline
        \end{tabular}
        \tablefoot{In all the spectra, a binning of a minimum of $50$ counts per bin has been used, with the only exception of the ASCA observation, for which the data were already reduced. A simplified identification code (ID) was assigned to each observation. Observations~2, 3, 4, and 5 are affected by high soft proton flaring background \citep[e.g.,][]{deluca04,marelli17}. Such high background, especially for Obs.~3, is responsible for the low number of counts for these observations, resulting in the high errors of the fit results of these observations. For this reason, we merged the spectra of Obs.~1, 2, and 3, Obs.~4 and 5, and Obs.~8 and 9. The MOS exposure times and counts are intended  to be the sum of such quantities for each camera. * ASCA counts are computed in the $0.5-10$ keV band.}
                       
\end{table*}
}
\newpage

\section{Tables of best-fit results}
\label{app:tables}
{\renewcommand{\arraystretch}{1.5}
\begin{table*}[h!]
\caption{Parameters of Abs. 1.}
\label{tab:abs1}
\centering
\begin{tabular}{lcccc}
                \hline
                ID & $N_{\text{H},1}/10^{21}$ cm$^{-2}$ & $\log(\xi_1/$erg cm s$^{-1}$) & $\Delta\chi^2$\\
                \hline
A & $7.5^{*}$ & $0.35^*$ & $275$\\
0 & $7.7_{-0.4}^{+0.3}$ & $0.34_{-0.01}^{+0.01}$ & $1990$\\
1+2+3 & $7.5^{*}$ & $0.35^{*}$ & $1377$\\
4+5 & $6.9_{-3.1}^{+0.7}$ & $\leq0.32$ & $1117$\\
6 & $7.5_{-0.5}^{+0.6}$ & $0.35_{-0.02}^{+0.04}$ & $1852$\\
7 & $7.9_{-0.3}^{+0.5}$ & $0.33_{-0.01}^{+0.01}$ & $1073$\\
8+9 & $7.4_{-0.5}^{+0.5}$ & $0.34_{-0.01}^{+0.02}$ & $2182$\\
10  & $7.6_{-1.0}^{+0.6}$ & $0.37_{-0.03}^{+0.06}$ & $880$\\
11  & $7.5^{*}$ & $0.35^{*}$ & $648$\\
                \hline
        \end{tabular}
                \tablefoot{We only show column density $N_\text{H}$ and ionization parameter $\log\xi$, since the velocity is not resolvable by the current X-ray dataset. We also show the significance of the absorber, $\Delta\chi^2$, with respect to the model with no absorbers, with $\Delta\text{dof}=2$ . All the null probabilities $P_\text{null}$ are less than $10^{-12}$ and therefore are not reported. $^{*}$For these observations, the two soft X-ray absorbers are not resolved, and therefore we fixed the parameters to the median values found for the other observations.}
\end{table*}
}

{\renewcommand{\arraystretch}{1.5}
\begin{table*}[h!]
\caption{Parameters of Abs.~2.}
\label{tab:abs2}
\centering
\begin{tabular}{lcccccc} % four columns, alignment for each
                \hline
                ID & $N_{\text{H},2}/10^{21}$ cm$^{-2}$ & $\log(\xi_2/$erg cm s$^{-1}$) & $z_{\text{o},2}/10^{-2}$ & $v_{2}/c$ & $\Delta\chi^2$ & $P_{\text{null}}$\\
                \hline
A & $7.8_{-1.4}^{+0.6}$ & $0.34_{-0.06}^{+0.63}$ & $+3.3_{-2.4}^{+3.9}$ & $0.102_{-0.037}^{+0.023}$ & $46$ & $10^{-12}$\\
0 & $1.8_{-0.1}^{+0.2}$ & $0.65_{-0.04}^{+0.10}$ & $+1.7_{-1.3}^{+1.4}$ & $0.117_{-0.012}^{+0.013}$ & $68.4$ & $10^{-11}$\\
1+2+3 & $3.7_{-0.1}^{+0.1}$ & $0.26_{-0.03}^{+0.05}$ & $-3.9_{-1.8}^{+1.2}$ & $0.172_{-0.018}^{+0.012}$ & $18.1$ & $10^{-3}$\\
4+5 & $7.7_{-0.2}^{+0.3}$ & $0.38_{-0.01}^{+0.01}$ & $+6.4_{-1.4}^{+1.5}$ & $0.073_{-0.013}^{+0.014}$ & $20.4$ & $3\times10^{-4}$\\
6 & $2.7_{-0.2}^{+0.2}$ & $0.06_{-0.02}^{+0.02}$ & $-2.1_{-1.4}^{+1.2}$ & $0.155_{-0.014}^{+0.012}$ & $61$ & $3\times10^{-11}$\\
7 & $6.7_{-1.0}^{+0.9}$ & $1.39_{-0.10}^{+0.07}$ & $+6.4_{-1.6}^{+1.3}$ & $0.073_{-0.015}^{+0.012}$ & $26.0$ & $6\times10^{-6}$\\
8+9 & $3.0_{-0.1}^{+0.2}$ & $0.63_{-0.03}^{+0.06}$ & $+1.1_{-0.8}^{+1.0}$ & $0.123_{-0.008}^{+0.010}$ & $73.4$ & $4\times10^{-15}$\\
10 & $3.0_{-0.2}^{+0.2}$ & $0.35_{-0.05}^{+0.04}$ & $+1.4_{-1.6}^{+1.2}$ & $0.132_{-0.016}^{+0.012}$ & $32.1$ & $10^{-6}$\\
11 & $5.4_{-0.6}^{+0.7}$ & $1.22_{-0.08}^{+0.07}$ & $+3.6_{-2.0}^{+1.0}$ & $0.100_{-0.019}^{+0.009}$ & $78$ & $3\times10^{-12}$\\
                \hline
        \end{tabular}
                \tablefoot{The column density $N_\text{H}$, ionization parameter $\log\xi$ , and observed redshift of the absorber $z_o$. The velocity of the absorber, in units of $v/c$, with $c$ the speed of light, is also shown. We report the significance of this absorber, $\Delta\chi^2$, with $\Delta\text{dof}=3$, and the null probability $P_\text{null}$ related to the model with one absorber.}
\end{table*}
}

{\renewcommand{\arraystretch}{1.5}
\begin{table*}[h!]
\caption{Parameters of Abs.~3.} 
\label{tab:abs3}
\centering
\begin{tabular}{lcccccccc}
                \hline
                ID & $N_{\text{H},3}/10^{21}$ cm$^{-2}$ & $\log(\xi_3/$erg cm s$^{-1}$) & $z_{\text{o},3}/10^{-2}$ & $v_{3}/c$ & $\Delta\chi^2_\text{2abs}$ & $P_{\text{null,2abs}}$ & $\Delta\chi^2_\text{1abs}$ & $P_{\text{null,1abs}}$\\
                \hline
A & $\geq440$ & $3.83_{-0.21}^{+0.21}$ & $+0.2_{-0.5}^{+0.5}$ & $0.132_{-0.005}^{+0.005}$ & / & / & / & /\\
1+2+3 & $143_{-69}^{+606}$ & $3.85_{-0.14}^{+0.47}$ & $+5.9_{-0.9}^{+0.5}$ & $0.078_{-0.008}^{+0.005}$ & $20.0$ & $4\times10^{-4}$ & $21.1$ & $3\times10^{-4}$\\
6 & $\geq295$ & $4.43_{-0.34}^{+0.22}$ & $-1.2_{-0.4}^{+0.5}$ & $0.145_{-0.004}^{+0.005}$ & $11.8$ & $0.01$ & $13.5$ & $0.01$\\ 
8+9 & $75_{-36}^{+125}$ & $3.77_{-0.13}^{+0.23}$ & $-4.7_{-0.7}^{+0.5}$ & $0.173_{-0.007}^{+0.005}$ & $12.0$ & $7\times10^{-3}$& $12.4$ & $0.01$\\
                \hline
        \end{tabular}
                \tablefoot{Again, column density $N_\text{H}$, ionization parameter $\log\xi$ , and the observed redshift $z_o$ are shown. For the ASCA observation and XMM-Newton Obs.~6, we are only able to report a lower limit for the column density. The velocity of this absorber is also reported in units of $v/c$. We show the significance of this absorber, $\Delta\chi^2_\text{2abs}$, with $\Delta\text{dof}=3$, and the null probability $P_\text{null,2abs}$, with respect to the model with two absorbers. Moreover, the same quantities with respect to the model with only Abs.~1, i.e. $\Delta\chi^2_\text{1abs}$ and $P_\text{null,1abs}$, are also shown. The UFO in the ASCA observation is not detected above a threshold confidence level of $2\sigma$. However, we report the best-fit result, in agreement with the previous claim by \cite{george97}.}
\end{table*}
}
{\renewcommand{\arraystretch}{1.5}
\begin{table*}[h!]
\caption{Blackbody temperatures and photon indices of each observation.} 
\label{tab:photonkT}
\centering
\begin{tabular}{lcccc}
                \hline
                ID & $kT/10^{-2}$ keV & $\Gamma$ & $\chi^2/\text{dof}$ & Fitted absorbers\\
                \hline
A & $6.4_{-1.1}^{+1.2}$ & $1.84_{-0.08}^{+0.05}$ & $538.0/427$ & 3\\
0 & $6.0_{-0.3}^{+0.2}$ &$1.68_{-0.02}^{+0.04}$ & $525.7/445$ & 2\\
1+2+3 & $7.0_{-0.2}^{+0.2}$ & $1.44_{-0.02}^{+0.04}$ & $500.3/464$ & 3\\
4+5 & $6.1_{-0.7}^{+0.6}$ & $1.68_{-0.06}^{+0.05}$ & $295.4/296$ & 2\\
6 & $5.5_{-0.3}^{+0.3}$ & $1.62_{-0.03}^{+0.03}$ & $389.2/372$ & 3\\
7 & $5.5_{-0.3}^{+0.6}$ & $1.63_{-0.07}^{+0.04}$ & $264.0/288$ & 2\\
8+9 & $6.3_{-0.4}^{+0.3}$ & $1.53_{-0.02}^{+0.02}$ & $649.4/663$ & 3\\
10 & $6.7_{-0.6}^{+0.9}$ & $1.58_{-0.06}^{+0.05}$ & $246.1/246$ & 2\\
11 & $6.7_{-0.3}^{+0.2}$ & $1.59_{-0.06}^{+0.09}$ & $232.0/195$ & 2\\
                \hline
        \end{tabular}
                \tablefoot{The final $\chi^2$ for the best-fit model of each observation is shown, as well as the number of absorbers fitted for each spectrum (see Tables~\ref{tab:abs1},~\ref{tab:abs2}, and \ref{tab:abs3} for details).}
\end{table*}
}

\newpage

\section{Comparison between XMM-Newton cameras}
\label{app:compa}

\begin{figure*}[h!]
\centering
\includegraphics[scale=0.5]{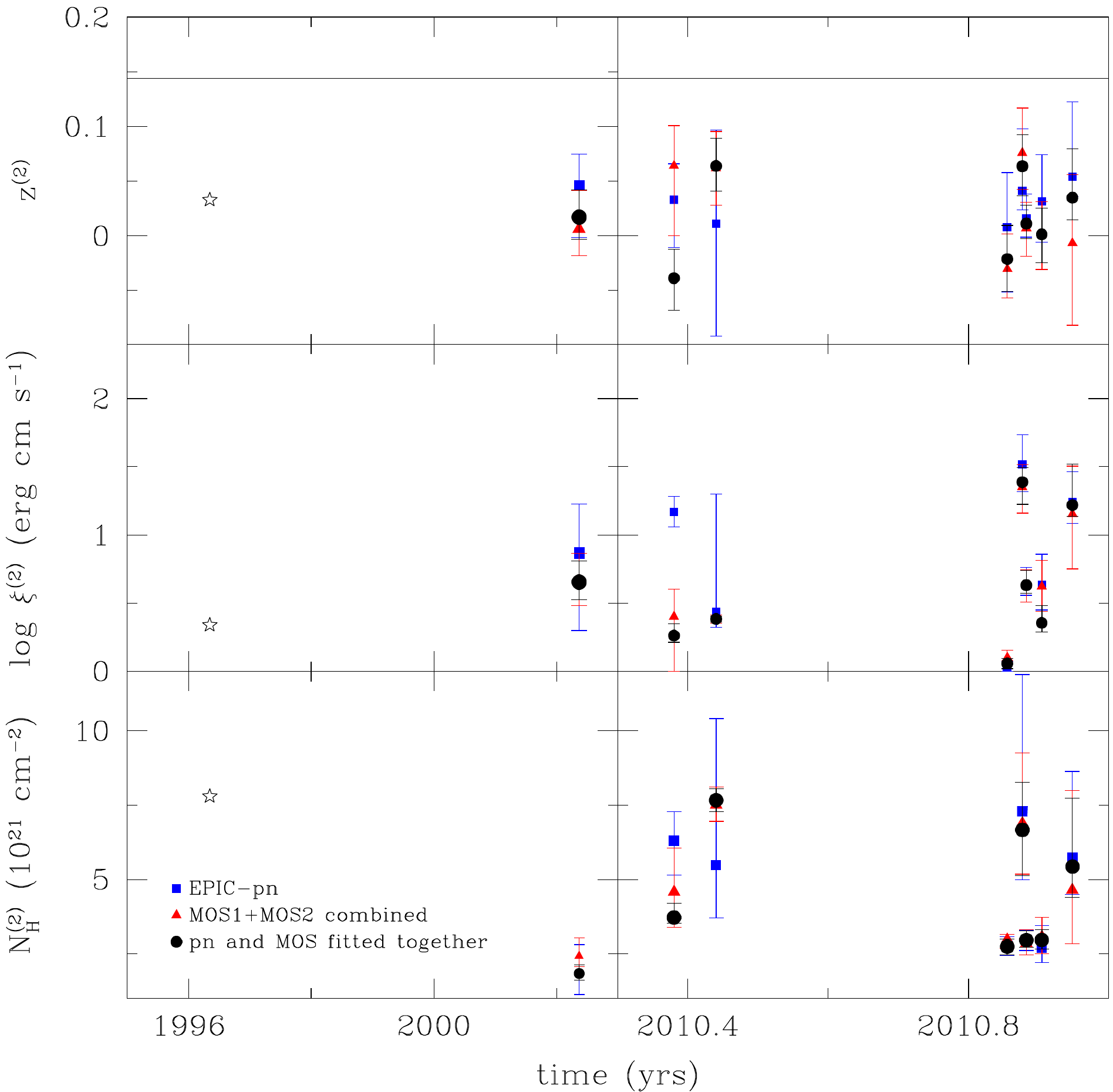}
\caption{Best-fit values of the pn camera (blue squares), the MOS cameras combined (red triangles), and pn and MOS data together (black circles) of the three parameters of Abs.~2. The star is the ASCA value. The parameters are in agreement within their errors at $90\%$ confidence level. The horizontal black line represents the cosmological redshift of the source, $z_o=0.144$.}
\label{fig:abs2}
\end{figure*}

\begin{figure*}[h!]
\centering
\includegraphics[scale=0.5]{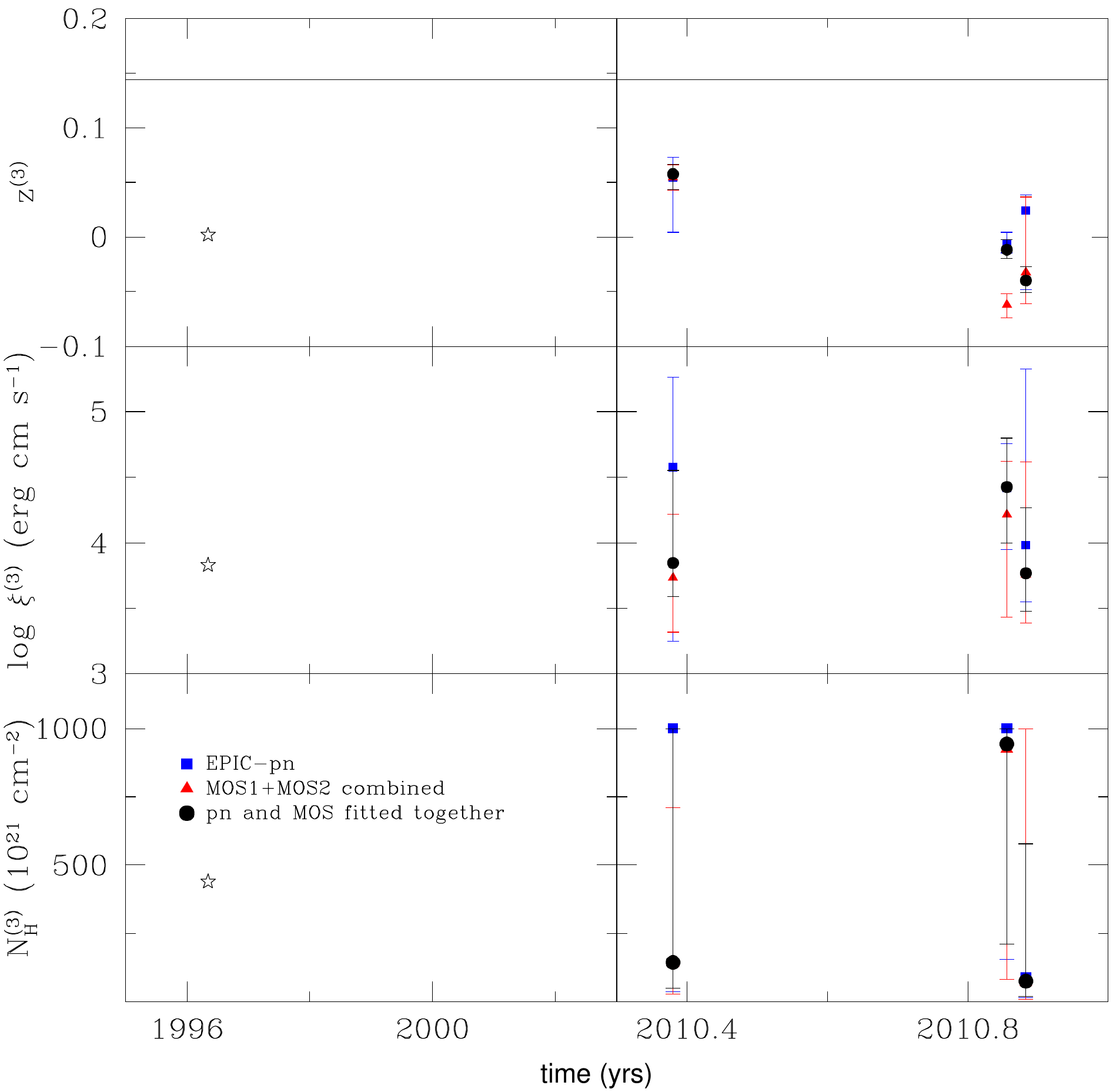}
\caption{Best-fit values of the pn camera (blue squares), the MOS cameras combined (red triangles), and pn and MOS data together (black circles) of the three parameters of Abs.~3. The star is the ASCA value. The parameters are in agreement within their errors at $90\%$ confidence level. The horizontal black line represents the cosmological redshift of the source, $z_o=0.144$.}
\label{fig:abs3}
\end{figure*}

\newpage

\twocolumn
\section{Spectra}
\label{app:spec}

\begin{figure}[h!]
\centering
\includegraphics[scale=0.35]{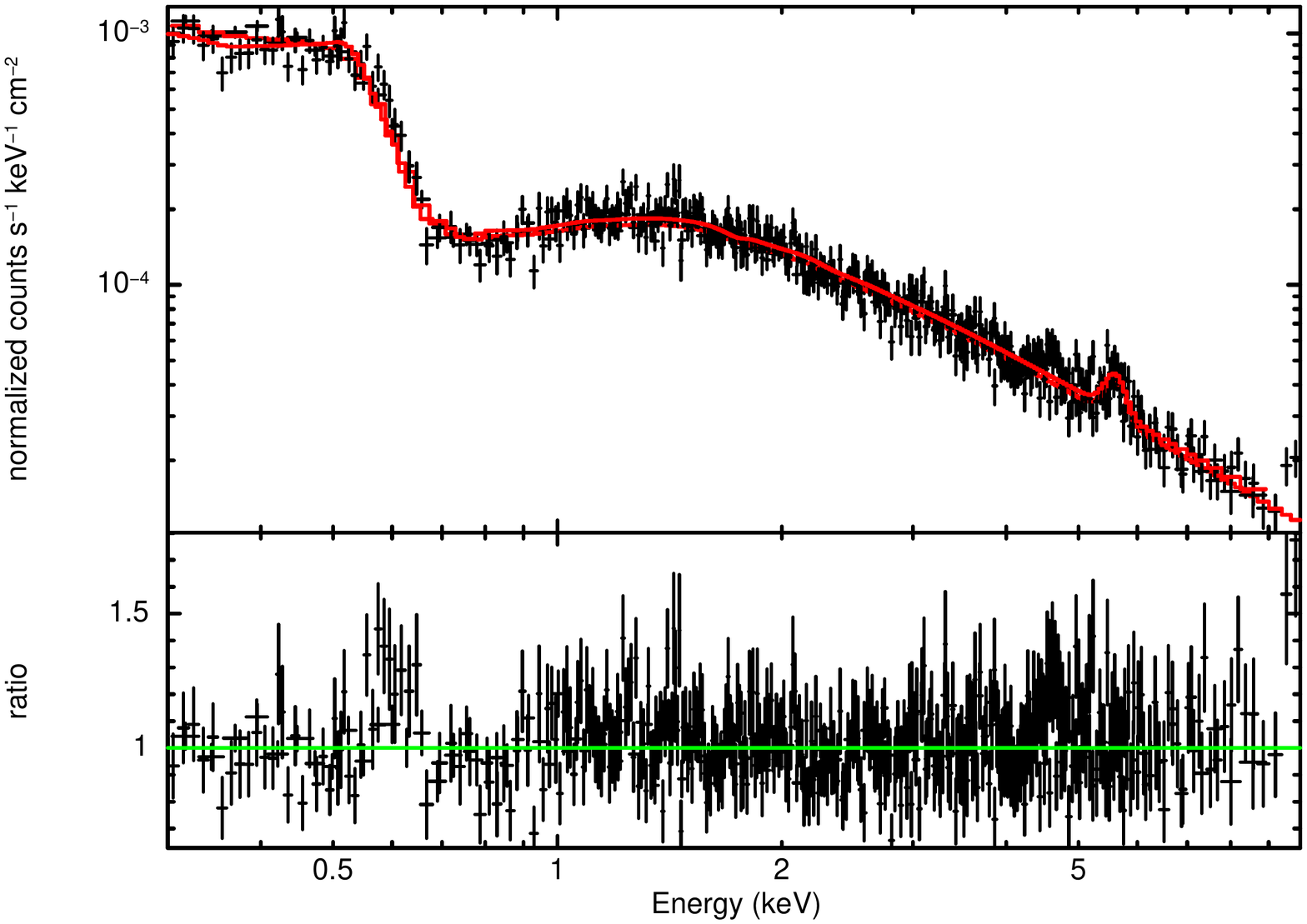}
\caption{Spectrum and ratio of Obs.~0. The red line is the best fit.}
\label{fig:obs00}
\end{figure}

\begin{figure}[h!]
\centering
\includegraphics[scale=0.35]{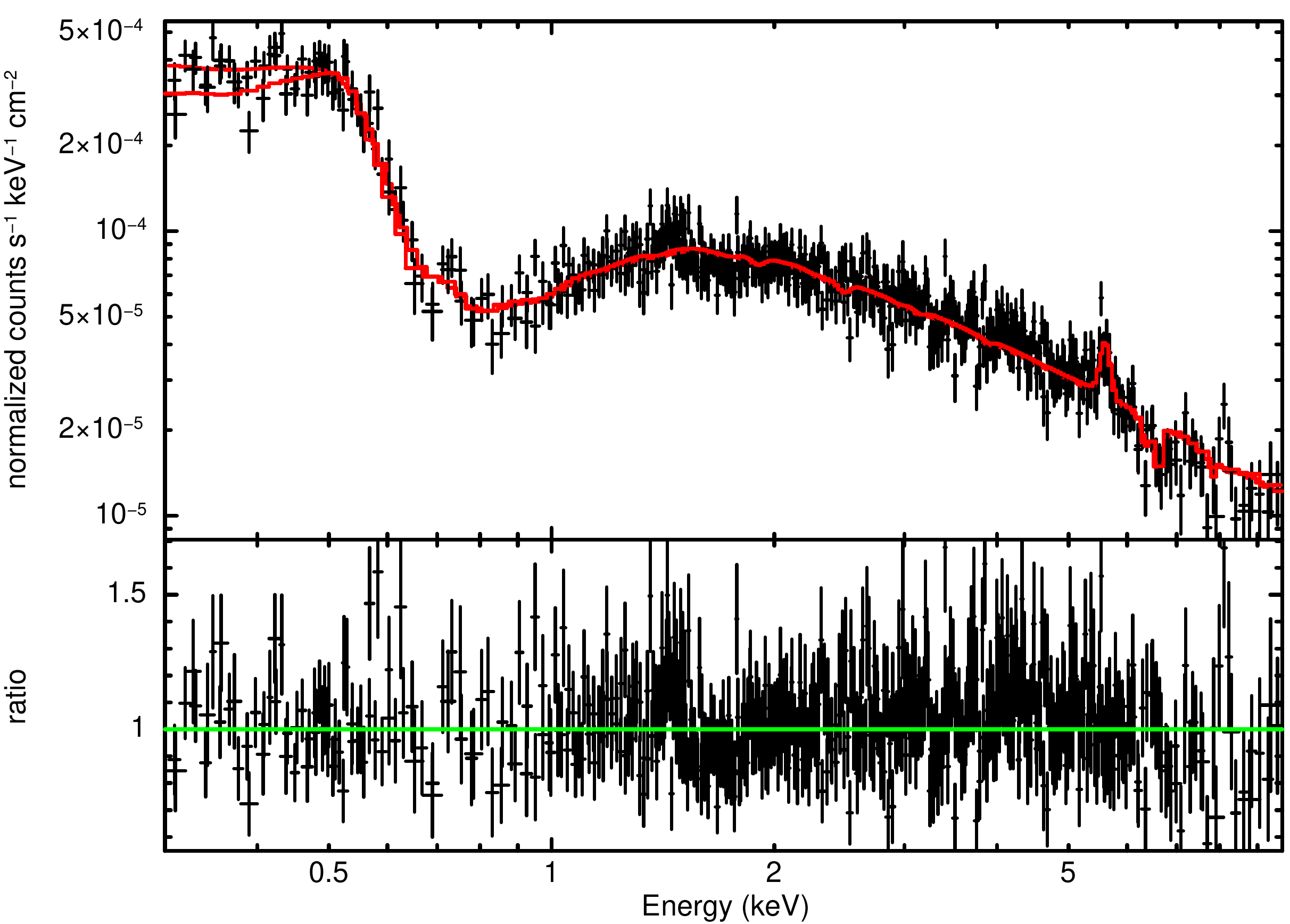}
\caption{Spectrum and ratio of Obs.~1+2+3. The red line is the best fit.}
\label{fig:obs010203}
\end{figure}

\begin{figure}[h!]
\centering
\includegraphics[scale=0.35]{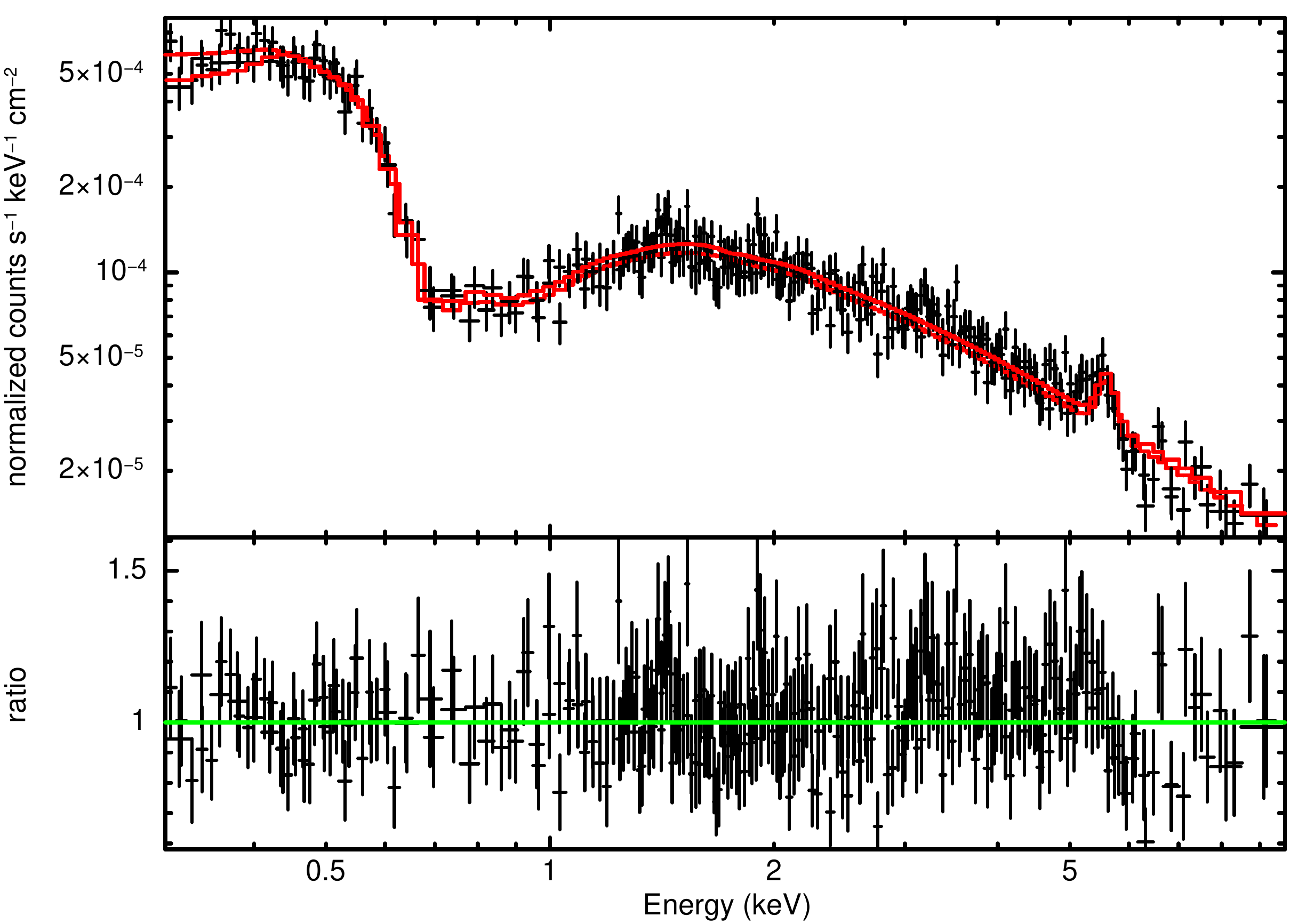}
\caption{Spectrum and ratio of Obs.~4+5. The red line is the best fit.}
\label{fig:obs0405}
\end{figure}

\begin{figure}[h!]
\centering
\includegraphics[scale=0.35]{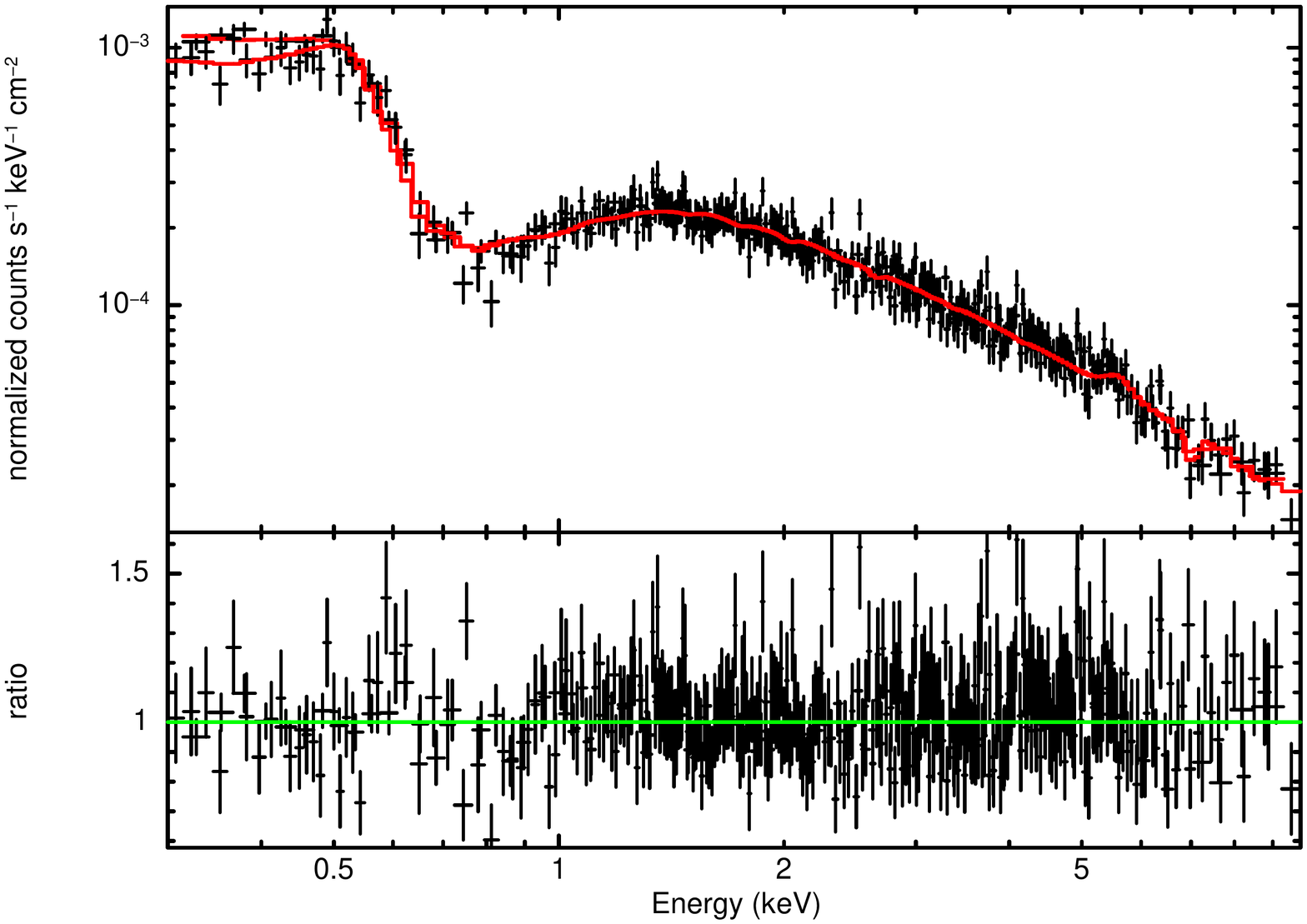}
\caption{Spectrum and ratio of Obs.~6. The red line is the best fit.}
\label{fig:obs06}
\end{figure}

\begin{figure}[h!]
\centering
\includegraphics[scale=0.35]{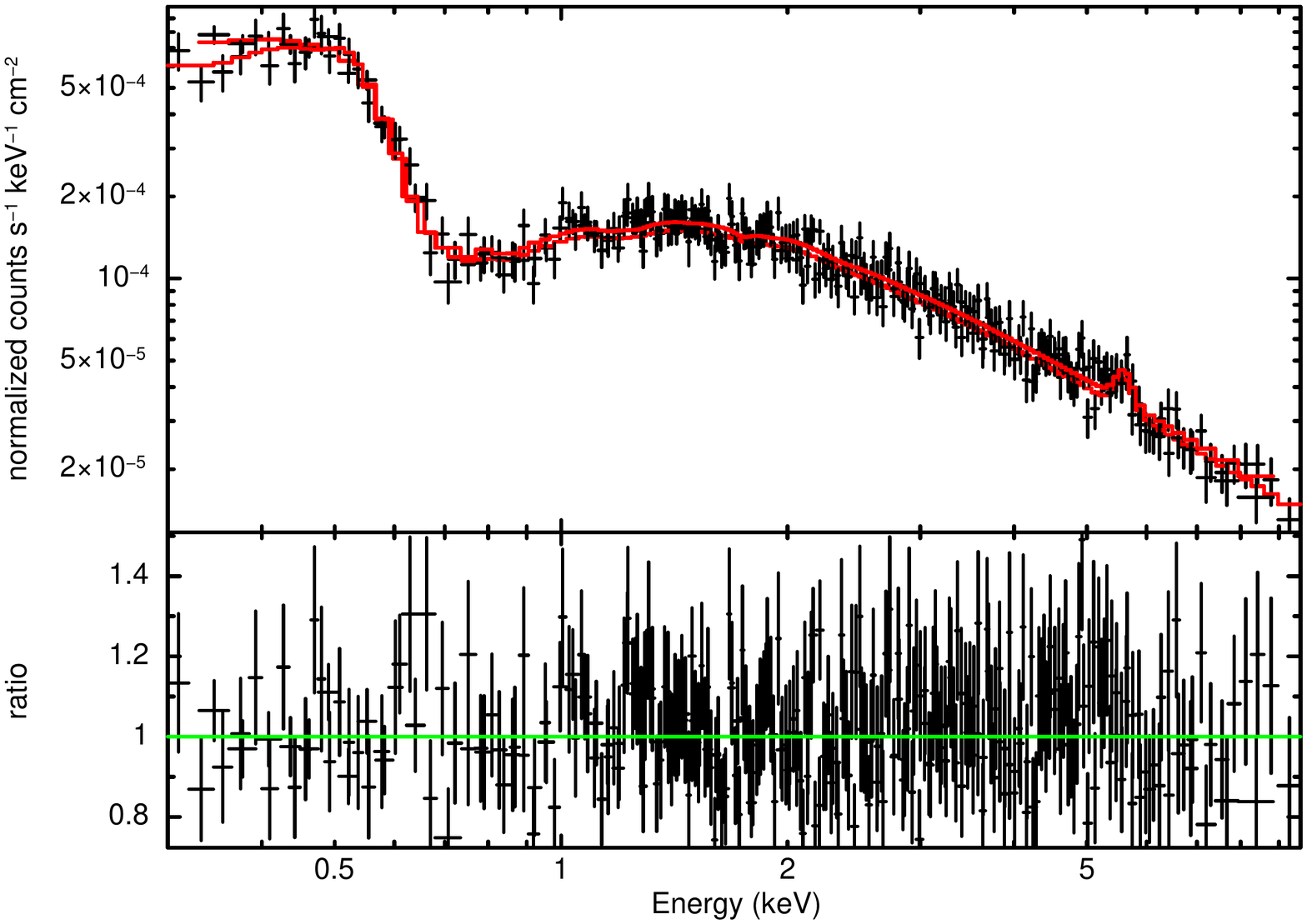}
\caption{Spectrum and ratio of Obs.~7. The red line is the best fit.}
\label{fig:obs07}
\end{figure}

\begin{figure}[h!]
\centering
\includegraphics[scale=0.35]{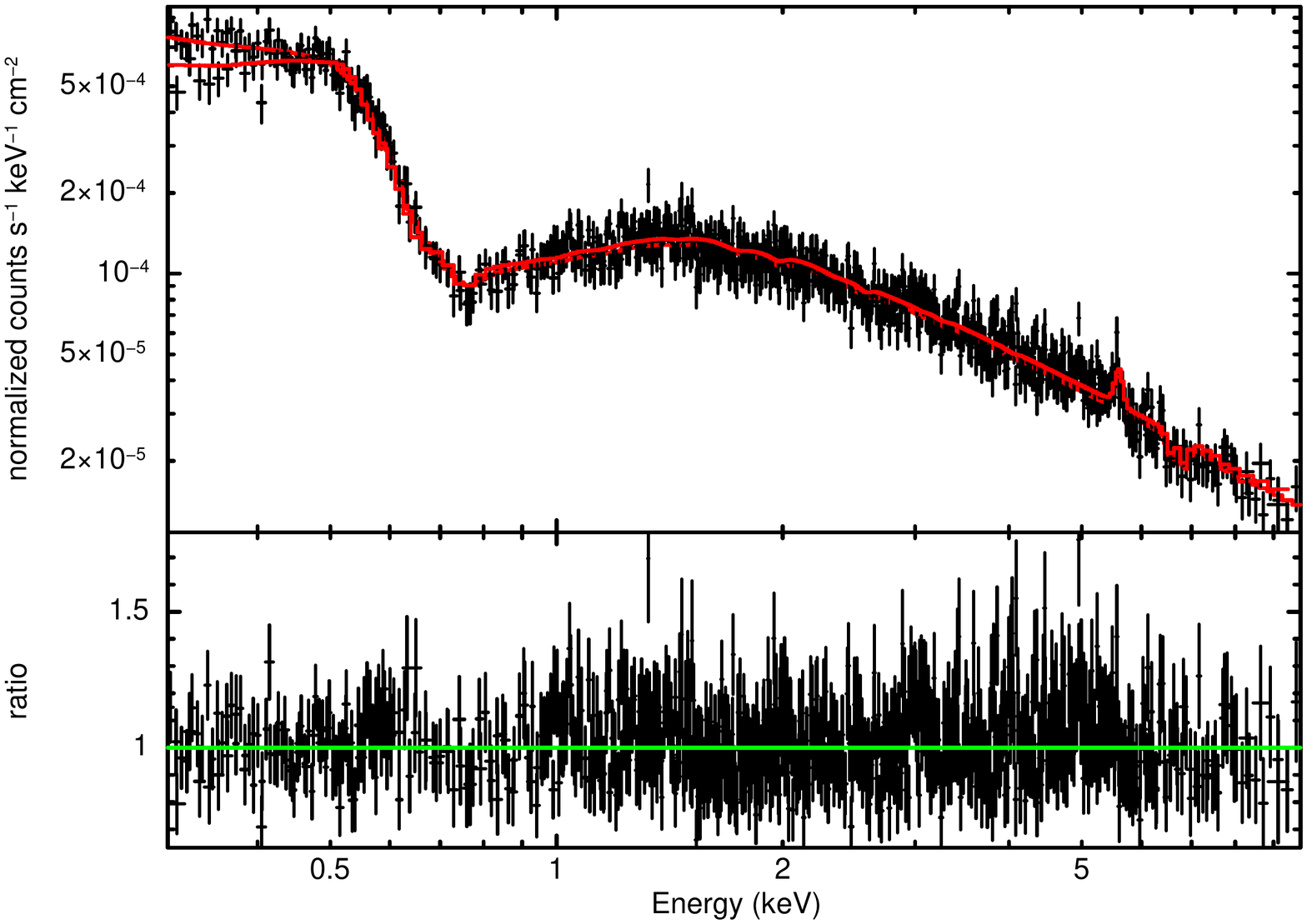}
\caption{Spectrum and ratio of Obs.~8+9. The red line is the best fit.}
\label{fig:obs0809}
\end{figure}

\begin{figure}[h!]
\centering
\includegraphics[scale=0.35]{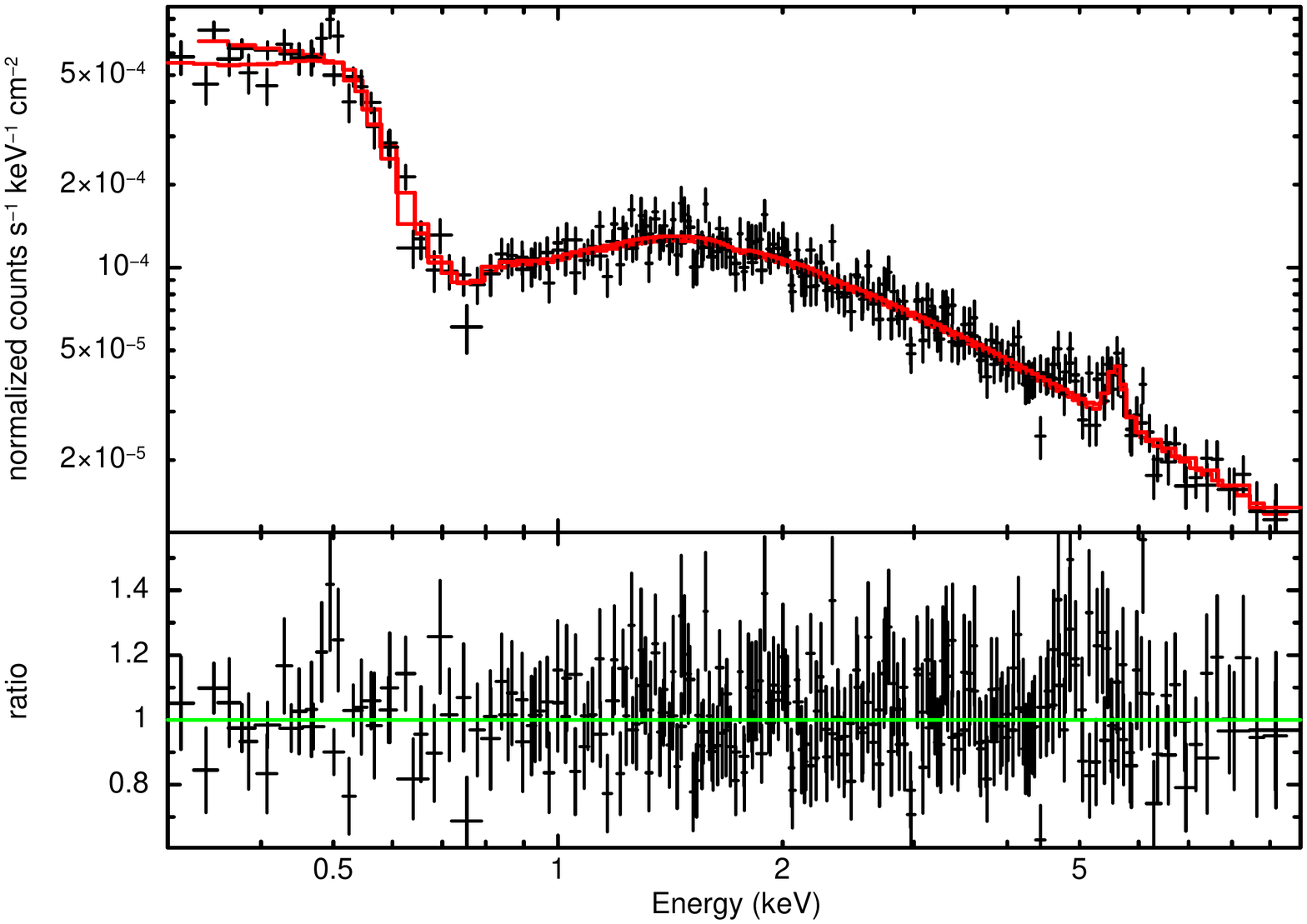}
\caption{Spectrum and ratio of Obs.~10. The red line is the best fit.}
\label{fig:obs10}
\end{figure}

\begin{figure}[h!]
\centering
\includegraphics[scale=0.35]{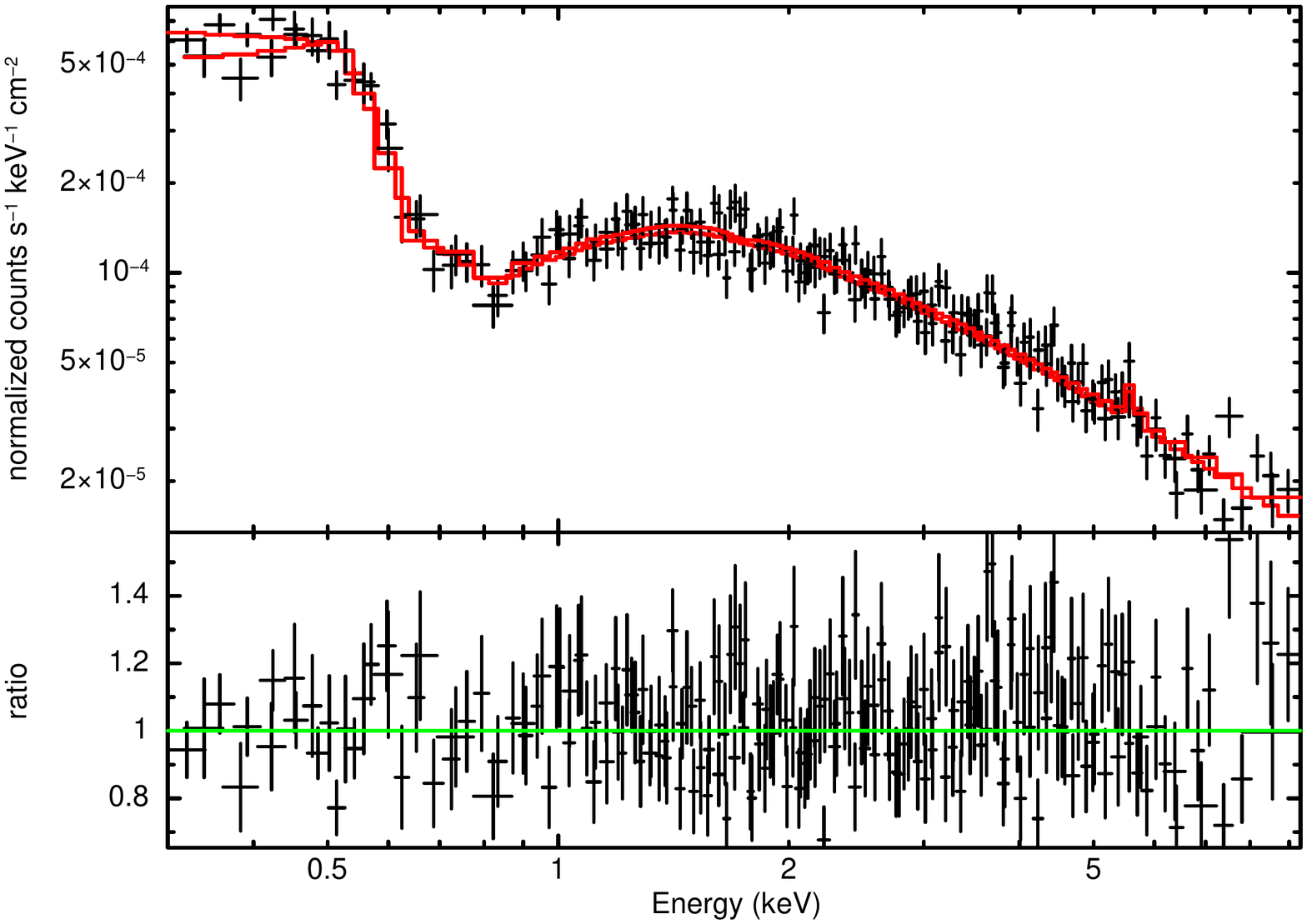}
\caption{Spectrum and ratio of Obs.~11. The red line is the best fit.}
\label{fig:obs11}
\end{figure}

\end{appendix}
%\begin{thebibliography}{}
%
%  \bibitem[Baker(1966)]{baker} Baker, N. 1966,
%      in Stellar Evolution,
%      ed.\ R. F. Stein,\& A. G. W. Cameron
%      (Plenum, New York) 333
%
%   \bibitem[Balluch(1988)]{balluch} Balluch, M. 1988,
%      A\&A, 200, 58

%   \bibitem[Cox(1980)]{cox} Cox, J. P. 1980,
%      Theory of Stellar Pulsation
%      (Princeton University Press, Princeton) 16
%   \bibitem[Cox(1969)]{cox69} Cox, A. N.,\& Stewart, J. N. 1969,
%      Academia Nauk, Scientific Information 15, 1
%
 %  \bibitem[Mizuno(1980)]{mizuno} Mizuno H. 1980,
  %    Prog. Theor. Phys., 64, 544
%   
 
%  \bibitem[Tscharnuter(1987)]{tscharnuter} Tscharnuter W. M. 1987,
 %     A\&A, 188, 55

  % \bibitem[Terlevich(1992)]{terlevich} Terlevich, R. 1992, in ASP Conf. Ser. 31, 
     % Relationships between Active Galactic Nuclei and Starburst Galaxies, 
%      ed. A. V. Filippenko, 13

  % \bibitem[Yorke(1980a)]{yorke80a} Yorke, H. W. 1980a,
   %   A\&A, 86, 286

   %%\bibitem[Zheng(1997)]{zheng} Zheng, W., Davidsen, A. F., Tytler, D. \& Kriss, G. A.
     % 1997, preprint
%\end{thebibliography}

\end{document}